\documentclass[fleqn,usenatbib]{mnras}

\usepackage{float}

\usepackage[T1]{fontenc}
\usepackage[utf8]{inputenc}
\usepackage{float}
\usepackage{url}
\usepackage{multirow}
\usepackage{widetext}
\usepackage{ae,aecompl}

\usepackage[dvipsnames]{xcolor}

\usepackage{graphicx}	
\usepackage{amsmath}	
\usepackage{amssymb}	
\usepackage{comment}
\usepackage{soul}

\usepackage{color}
\newcommand{\FS}[1]{\textcolor{black}{ #1}}
\definecolor{jm}{rgb}{0.23, 0.53, 0.75}

\definecolor{np}{RGB}{0, 255, 0}
\definecolor{js}{RGB}{0, 0, 0}
\definecolor{red}{RGB}{255, 0, 0}

\title[Extended PBH mass functions with a spike]{Extended primordial black hole mass functions with a spike}
\author[Magaña et al]{J. Maga\~na$^{1}$\thanks{juan.magana@ucentral.cl}, M. San Mart\' in$^{2}$, J. Sureda$^{3}$, M. Rubio$^{4,5}$, I. Araya$^{6,7}$ and N. Padilla$^{8}$
\\
$^{1}$Escuela de Ingenier\'ia, Universidad Central de Chile, Avenida Francisco de Aguirre 0405, 171-0164 La Serena, Coquimbo, Chile\\
$^{2}$Centro de Astro-Ingenier\'\i a, Pontificia Universidad Cat\'olica de Chile, Vicu\~na Mackenna 4860, Santiago, Chile\\
$^{3}$Donostia International Physics Center (DIPC), Paseo Manuel de Lardizabal, 4, 20018 Donostia-San Sebastián, Spain\\
$^{4}$SISSA, Via Bonomea 265, 34136 Trieste, Italy\\
$^{5}$Institute for Fundamental Physics of the Universe (IFPU), Via Beirut 2, 34014 Trieste, Italy\\
$^{6}$Instituto de Ciencias Exactas y Naturales (ICEN), Universidad Arturo Prat, Avenida Arturo Prat Chac\'on 2120, 1110939, Iquique, Chile\\
$^{7}$Facultad de Ciencias, Universidad Arturo Prat, Avenida Arturo Prat Chac\'on 2120, 1110939, Iquique, Chile\\
$^{8}$Instituto de Astronomía Teórica y Experimental (IATE), CONICET-UNC, Laprida 854, X5000BGR, Córdoba, Argentina\\
}

\date{Accepted XXX. Received YYY; in original form ZZZ}

\pubyear{2022}

\begin{document}
\label{firstpage}
\pagerange{\pageref{firstpage}--\pageref{lastpage}}
\maketitle

\begin{abstract}
We introduce a modification of the Press-Schechter formalism aimed to derive general mass functions for primordial black holes (PBHs). In this case, we start from primordial power spectra (PPS) which include a monochromatic spike, typical of ultra slow-roll inflation models. We consider the PBH formation as being associated to the amplitude of the spike on top of the linear energy density fluctuations, coming from a PPS with a blue index. By modelling the spike with a log-normal function, we study the properties of the resulting mass function spikes, and compare these to the underlying extended mass distributions. When the spike is at PBH masses which are much lower than the exponential cutoff of the extended distribution, very little mass density is held by the PBHs within the spike, and it is not ideal to apply the Press-Schechter formalism in this case as the resulting characteristic overdensity is too different from the threshold for collapse.  It is more appropriate to do so
when the spike mass is similar to, or larger than the cutoff mass.  Additionally, it can hold a similar mass density as the extended part. Such particular mass functions also contain large numbers of small PBHs, especially if stable PBH relics are considered, and they can provide $\sim 1000M_\odot$ seeds for the supermassive black holes at the centres of present-day galaxies. The constraints on the fraction of dark matter in PBHs for monochromatic mass functions are somewhat relaxed when there is an additional underlying extended distribution of masses.
\end{abstract}




\section{Introduction}

There are many reasons why the study of primordial black holes (PBHs) has attracted much of the community of cosmologists for the last almost 50 years. As it is well known, there is strong and highly convincing body of astrophysical evidence for the fact that more than 80\% of the matter in the Universe should be rather cold and non-baryonic. However, so far there is no concrete detection of this kind of matter, nor is there a clear consensus on how it is created, how it evolves or what it is composed of. PBHs constitute plausible candidates for explaining this matter content. The idea of PBH formation as a result of the collapse of density fluctuations in the Early Universe, rather than of stellar collapse, has been proposed in seminal works by \citet{Zeldovich1966, Hawking1971,Carr_Hawking:1974}. Depending on the mechanism by which they are formed, PBHs may be endowed with masses ranging from the Planck scale ($\sim 10^{-38} M_{\odot}$) to stupendously large scales ($\sim 10^{15} M_{\odot}$) \citep{Carr:2020_slargepbhs}. Nevertheless,  PBHs with masses on the order of that of one asteroid mass are among the most likely candidates, being able to constitute a sizeable fraction of the dark component of the Universe \citep{Coogan:2020tuf,Smyth:2019whb,Montero-Camacho:2019jte,Martin19,Ray:2021mxu}. Of course, the question still remains as to whether or not all dark matter is actually made of pure PBHs, for which observations will hopefully provide an answer in due course. 

While the evidence for PBHs is far from conclusive, there is a growing appreciation for the variety of potentially important roles in astrophysical and cosmological scenarios. The recent detection of gravitational waves from LIGO and Virgo \citep{LIGOScientific:2016vpg,LIGOScientific:2017vwq,LIGOScientific:2020,LIGOScientific:2021} has also contributed to the resurgence of interest in these objects as they could even explain the origin of the most massive coalescing sources. This hypothesis, which should be able to be tested in the near future (and therefore possibly ruled out), also gives \FS{PBHs the possibility of being seeds for SMBHs in galactic nuclei} \citep{Volonteri:2021,Papanikolaou:2020qtd}, since in principle there is no solid reason to rule out the possibility of PBHs to have maximum masses in the “supermassive” range, although this particular possibility may have already been ruled out \citep{Sureda:2021}. In fact, it is of particular interest the detection of a binary system in which one of its components has a mass in the gap between the heaviest neutron star and the lightest stellar black hole \citep{LIGOScientific:2020, LIGOScientific:2021}, since the origin of the latter could be likely primordial. On the other hand, although in tension with LIGO/Virgo measurments, some authors claim that PBH formation at inflationary epochs could be responsible for the stochastic gravitational wave signal evidence in the 12.5 yrs pulsar timing array data by  NANOGrav \citep{DeLuca:2021prl, Vaskonen:2021prl}. Likewise, a feature that appeals to the relativistic physics community is PBH evaporation, particularly of those with low mass, and their consequent gamma-ray radiation, which could provide a test of critical phenomena at these scales. PBHs of masses less than 1 kg have already evaporated and therefore have not survived to the present day. However, this phenomenon should have influenced several physical processes that occurred in the Early Universe, and therefore provide tests to verify, or not, their existence. 

Among the main scenarios and mechanisms for the formation of PBHs, is the collapse of highly dense (adiabatic) perturbations, which were generated in an inflationary period of the Early Universe. This has involved a series of advances towards a better understanding of the threshold for black hole formation, as well as its dependence on the type and shape of such a perturbation. In general, in order to give a rather standard estimate of the abundance of PBHs and its corresponding mass function, it is assumed that these perturbations follow a Gaussian probability distribution, and that the shape of the peak strongly depends on the primordial power spectrum. However, it is known that this is no longer suitable when the perturbation spectra exceed a maximum value of overdensity. Other possible formation mechanisms could be first and second-order phase transitions \citep{Jedamzik:1999PhRvD,Rubin:2000hep,Liu:2022PhRvD,Kawana:2022PhLB}, non-linear metric perturbations \citep{Allahyari:2017JCAP,Hidalgo:2009PhRvD}, enhancement in the power spectrum of density fluctuations \citep{Cole:2022arXiv220407573C,Ballesteros:2021fsp,Ashoorioon:2019xqc}, collapse of cosmic strings \citep{Polnarev:1991,Wichoski:1998ev}, single and multifield inflation scenarios \citep{Ahmed:2022NuPhB,Ragavendra:2021PhRvD,Palma:2020ejf}, among others.

A lot of work has been invested in inferring the mass function of PBHs. A first approximation considers a \textit{monochromatic} mass function, assuming that all the PBHs have the same mass. This assumption is actually followed by the different observational constraints on the fraction $f$ of PBHs as dark matter, coming from strong lensing effects by PBHs \citep{Tisserand:2007,Ogle:2011}, \FS{microlensing effects \citep{Niikura17Nature,Niikura19PRD,Blaineau22}}, dynamical friction, accretion disks, large-scale structure formation, PBH evaporation contribution to the gamma-ray background, the $511$ keV emission detected at the center of our galaxy \citep{Laha:2019PhRvL.123y1101L,Laha:2020PhRvD.101l3514L}, and  nucleosynthesis processes, among others \citep[see][for a review of observational constraints]{Carr:2020_constraints}. When combining all these constraints, there remains  a small range of masses where the dark matter could be composed fully by PBHs ($f_{\text{PBH}}=1$). Nonetheless, such constraints could change considerably under the hypothesis that PBHs span  a wide range of masses. A seminal work by \citet{Carr:1975} argued that the mass function for PBHs formed from scale-invariant inhomogeneities at early cosmic stages follows a \textit{power-law}, whose exponent depends on the equation of state of the Universe at the formation time. Also, an interesting mass spectrum for PBH formation is suggested by \citet{Dolgov:1993PhRvD} from a baryogenesis mechanism based on spontaneous charge symmetry breaking. In this last case the mass distribution turns out to be a symmetric \textit{log-normal} with a peak at an associated mass scale \citep[see also][]{Garcia-Bellido:1996PhRvD,Clesse:2015PhRvD}. \FS{Consequences on the abundance of PBHs from the form of the power spectrum and the role of non-gaussianities have been recently explored \citep{Germani:2018jgr,Germani:2019zez,Atal:2018neu}.}

The scientific interest in studying extended mass functions grew even more from  observations by Cosmic Microwave Background (CMB) radiation satellites, including the Planck Satellite, suggesting that the primordial power spectrum is almost \textit{scale-invariant}, leading to the need for enhancing it in order to predict a significant population of PBHs. To this end,  \citet{Sureda:2021} proposed a modified Press-Schechter formalism for making full explicit calculations of the extended PBH mass distributions from a broken power-law primordial spectrum. The variance of the fluctuations is computed assuming two different formation scenarios: (i) \textit{fixed conformal time}, in which the formation for PBHs of different masses occurs at the end of the inflationary epoch with the same scale factor, and (ii) \textit{horizon crossing}, where PBH formation occurs when the fluctuations enter into causal contact within the cosmological horizon (depending thus on the PBH mass, being the lighter PBHs formed earlier). The authors reported three different regions in the parameter space where all the DM could consist entirely of PBHs and, interestingly, within the range of masses coming from LIGO/Virgo measurements. After that, \citet{Padilla:2021} presented a detailed study on how the non-linear power spectrum, the abundance of haloes and their clustering are modified because of Poisson noise due to PBH discreteness, when they obey the extended mass functions constructed by \citet{Sureda:2021}. By demanding (at most) 10\% of deviations from the standard model predictions, their constraints are in agreement to those obtained by \citet{Sureda:2021}, pointing out that these windows could be consistent with LIGO/Virgo detections corresponding to PBHs.

The possibility of making full explicit calculations of the PBH mass functions naturally suggests considering the mass spectrum to contain one or more spikes. Indeed, there seems to be a strong relation between the shape of the inflationary model and the resulting primordial power spectrum, for which some bumpy scalar potentials for the inflaton could imprint a peak at a certain scale, producing PBHs in a rather narrow range of masses. Likewise, some other inflationary potentials predict double (or even multiple) peaks in the power spectrum, thus implying a double-peaked (or multiply-peaked) PBH mass function. Also, phase transitions at the inflationary stage could provide spikes in the spectrum of fluctuations at any scale. A recent proposal by \citet{Carr:2019} is to choose the non-Bunch-Davies vacua, which involve oscillations in the inflationary power spectrum that generates oscillations in the mass functions with well-pronounced spikes. A new way to produce extended mass functions with a spike at any mass scale was proposed in \citet{Carr_Clesse:2021PDU}, reporting a steeper power spectrum for which the amplitude of the fluctuations that are greater than a certain \textit{pivot} wavenumber $k_{piv}$ are amplified. Moreover, the threshold density for collapse turns out to be a function of the equation of state, $w$, which depends on the thermal history of the Universe; i.e., of the temperature field. Given that there is a relation between the temperature of the Universe and the PBH mass at the formation time, the threshold density becomes a function of the PBH mass in turn. Thus, when the evolution of the threshold density is used in order to estimate the PBH abundance, the obtained extended mass distribution results in a spiky function \citep[see also][]{Papanikolaou:2022cvo}.

In this work, we extend the Press-Schechter extended mass functions  in \citet{Sureda:2021}, by endowing them with a log-normal spike. This is justified because, as described above, many inflationary scenarios and models for the physics of the very-early universe tend to create power-spectra with characteristic features of excess power at definite scales, which in turn lead to mass functions with spikes. Thus, we view the option of adding log-normal distributions to the primordial power-spectrum as a way of accounting for these features of excess power, leading to mass-functions which have features that are intermediate between a monochromatic distribution and the Press-Schechter extended distribution.

This paper is organized as follows. In Section \ref{sec:PPS}, we introduce a primordial power spectrum for density fluctuations composed by broken power law with a spike feature. In Section \ref{sec:broad_mass_function_formalism}, we 
propose a modification of the Press-Schechter formalism to derive an extended PBH mass function starting from the proposed primordial power spectrum. In Section \ref{sec:constraints}, we present the constraints on the fraction of dark matter as PBHs described by an extended mass distribution with a spike. 
Finally, in Section \ref{sec:conclusions}, we summarise our main results and conclusions.

For the numerical simulations throughout this work, we consider a flat cosmology with the following density parameters as measured by Planck Collaboration \citep{Planck:2018}: $\Omega_{m,0} = 0.315$, $\Omega_{dm,0} = 0.264$, $\Omega_{r,0} = 9.237\times 10^{-5}$, and a Hubble constant $H_{0}=67.36\, \mathrm{km \,s^{-1} Mpc^{-1}}$

\section{The primordial power spectrum}\label{sec:PPS}
In this section, we introduce a modification to the Press-Schechter formalism, in order to construct extended PBH mass functions endowed with a spike. For doing so, we propose a modified broken power spectrum for the density fluctuations including a spike, and follow the guidelines presented in \citet{Sureda:2021}. 


The simplest picture for PBH formation essentially consists of assuming that PBHs are originated from the collapse of density fluctuations present in the radiation-domination epoch of the Universe \citep{Khlopov_2010}. A perturbation in the density field can be characterised by its \textit{density contrast}
\begin{equation}
    \delta:=\frac{\rho_{f}-\rho_{b}}{\rho_b},
\end{equation}
where $\rho_{f}$ and $\rho_{b}$ are the fluctuating and background densities, respectively. If the fluctuation density contrast overcomes certain \textit{threshold density} $\delta_{c}$, a PBH is formed. Nevertheless, the size of typical perturbations in the early Universe is rather tiny, which means that PBH production becomes actually a rare event. Thus, if one is expected to get a significant abundance of PBHs, a different mechanism would be needed. One plausible way for triggering their formation is by enhancing the amplitude of the density fluctuations, allowing to them the capability to reach such a threshold density. Given that the amplitude of the fluctuations is described by the primordial power spectrum, it is this characteristic of the density field that should be modified to make the amplitude of fluctuations higher at a certain scale, thus accordingly enhancing the primordial power spectrum.

Recently, \cite{Sureda:2021} proposed a \textit{broken power law} for modelling the corresponding spectrum, in such a way that small-scale fluctuations follow a power-law with a  blue index $n_{b}>1$, whereas the corresponding power-law for large-scale fluctuations considers a red spectral tilt ($n_s<1$), as recently measured and reported by the Planck collaboration \citep{Planck:2018}. This broken power law power spectrum corresponds to the primordial spectrum of energy density fluctuations, and it is given by the following piece-wise function,
\begin{equation}
P_{\text{BPL}}(k) = \begin{cases}
A_1 k^{n_s} &\text{for $k<k_{\text{piv}}$},\\
A_1 k_{\text{piv}}^{n_s-n_b}\, k^{n_{b}} &\text{for $k\geq k_{\text{piv}}$},
\end{cases}
\label{eq:pkbroken}
\end{equation}

\noindent
where $k$ is the wavenumber associated to the scale of fluctuations, $A_{1}=A_s/k_{0}^{n_s}$, and $A_s = 2.101\times 10^{-9}$ are the normalization constants for the power spectrum measured at $k_{0}=0.05\, \mathrm{Mpc}^{-1}$. The tilt of the power spectrum becomes blue for wavenumbers greater than the \textit{pivot wavenumber}, $k_{\text{piv}}$. Although the choice of the $k_{\text{piv}}$ value is arbitrary,   CMB measurements from Planck and Atacama Cosmology Telescope \citep[ACT,][]{ACT:2020gnv} determine a nearly scale-invariant power spectrum with  $n_s\approx 1$ up to $k\sim \mathcal{O}(1)\, \mathrm{Mpc}^{-1}$. Therefore, the broken power law spectrum is fixed by appropriate parameters that enhance the small-scale fluctuations beyond these scales.


On the other hand, there are proposals for inflationary potentials that produce spikes in the primordial power spectrum \citep{Mishra:2019pzq,Palma:2020ejf,Ozsoy:2020kat,LIU:2021,Zheng_2022,Pi:2021arXiv211212680}. The collapse of density inhomogeneities described by such primordial power spectra could give origin to  broad distributions of PBH masses, with an additional population of PBHs with masses corresponding to the scale of the spike. If the spike is narrow, it can be ideally modelled with a delta function centered at the spike, namely $\delta(k-k_{\text{p}})$. Alternatively, it can be modeled with a \textit{log-normal} function,
\begin{equation}
P_{\text{LN}}(k, k_{\text{p}})=\frac{A_2 }{\sqrt{2\pi} \epsilon} \exp{\left[-\frac{\log^2{(k/k_\text{p})}}{2\epsilon^{2}}\right]},
\label{eq:pk_lognormal}
\end{equation}
where $A_{2}$ is the amplitude of the spike, $\epsilon$  its width, and $k_{\text{p}}$, the wavenumber of the peak. Notice that when $\epsilon \to 0$, this function resembles a Dirac delta function, which corresponds to a monochromatic distribution in power. We are interested in studying a smooth approximation of the power spectrum resulting from an extended component, and a monochromatic source. Also, whatever process produces such a strong enhancement in the power spectrum at a narrow range of scales, should be considered non-linear. We will ignore this consideration in the present paper, although there are several proposals for an improved treatment of enhanced fluctuations for the formation of PBHs \citep[see for instance][and references therein]{Masahiro:2019,Kalaja:2019JCAP,Young:2020xmk}.

\begin{figure}
  \centering
  \includegraphics[width=0.45\textwidth]{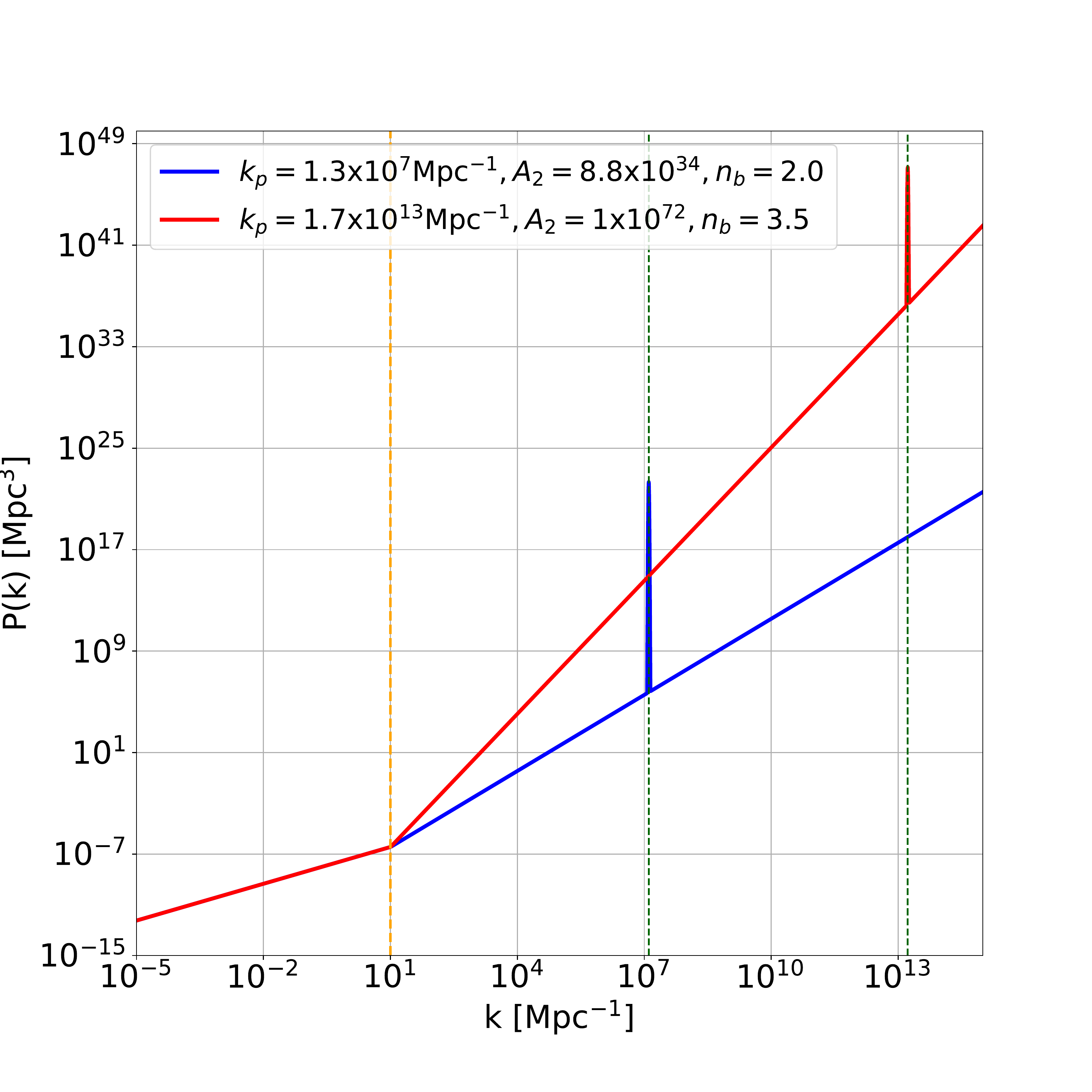}
\caption{Primordial power spectra as a function of the wavenumber $k$ (given by Eq. \eqref{eq:pk_plspike}) with $n_{s}=0.9649$, and broken power law blue indices $n_{b}=2$ (in blue) and $n_{b}=3.5$ (in red). We take a pivot wavenumber $k_{\text{piv}}=10\,\mathrm{Mpc}^{-1}$ (orange vertical line) and $\epsilon=0.01$. The gray vertical lines show two different scales for the spike wavenumber, $k_{\text{p}}$.
} 
\label{fig:powerspectrum}
\end{figure}

We propose the following primordial power spectrum, composed as a broken power-law with a spiky feature at a certain characteristic wavenumber, namely
\begin{equation}
P(k)= P_{\text{BPL}}(k)+ P_{\text{LN}}(k), 
\label{eq:pk_plspike}
\end{equation}
where $P_{\text{BPL}}$ is the broken power law spectrum given by Eq. \eqref{eq:pkbroken}, and the spiky feature is given by the log-normal function $P_{\text{LN}}(k)$ introduced in \eqref{eq:pk_lognormal}.
Figure \ref{fig:powerspectrum} shows two examples of this power spectrum for two different values for the blue index. Notice the presence of the \textit{log-normal} spike in power for wavenumbers $k_p$ above  the pivot scale. 

\section{Extended mass function with a spike}
\label{sec:broad_mass_function_formalism}

Having introduced an alternative primordial power spectrum, the statistical properties of the density field can be encoded in the averaged variance of the inhomogeneities. The variance depends on the primordial power spectrum as
\begin{equation} \label{eq:sigmasquared}
\sigma^{2}(M)  =4\pi D^2(a)\int_{0}^{\infty}{
k^{2}P(k)\widehat{W}^{2}(R(M),k)\,\mbox{d}k},
\end{equation}
where $M=4/3 \pi (aR(M))^3 \rho_b$ is the total energy  in a sphere of radius $aR$, with $\rho_b$ the background density, $k$ is the comoving wave-number in Fourier space, $P(k)$ is the corresponding primordial power spectrum, $ D(a)$ is the growth factor of the fluctuations  at a certain scale factor $a$, and $\widehat{W}$ is a window function with support $R(M)$.

Also, it is common to define the \textit{height} of the peak as
\begin{equation}\label{eq:nu}
\nu(M)=\frac{\delta_{c}}{\sigma(M)},
\end{equation}
where $\delta_{c}$ is the linear threshold density contrast for PBH formation and $\sigma(M)$ the variance defined in (\ref{eq:sigmasquared}).

We now construct the corresponding extended mass functions. To do so, we follow a modified version of the Press-Schechter formalism, for which the mass function reads
\begin{eqnarray}\label{eq:dndm_gen}
\left(\frac{\mbox{d}n}{\mbox{d}M}(M)\right)_{\tiny{\mbox{PS}}} &=&
\nu f(\nu)\frac{\rho_{\tiny{\mbox{DM}}}}{M^{2}}\frac{\mbox{d}\log\nu}{\mbox{d}\log M} \nonumber \\
&=& f(\nu)\frac{\rho_{\tiny{\mbox{DM}}}}{M}\frac{\mbox{d}\nu}{\mbox{d}M},
\end{eqnarray}
where $\rho_{\tiny{\mbox{DM}}}$ is the dark matter density, and $f(\nu)$ is the multiplicity function of the density field.

\subsection{General assumptions}

Following \citet{Sureda:2021}, we consider two scenarios of PBH formation, referred to as \textit{fixed conformal time} (FCT) and \textit{horizon crossing} (HC). In the former, all the PBHs covering a wide range of masses are formed at the same epoch. On the other hand, in the HC scenario, a PBH is formed when its linear fluctuation size matches within the horizon and its mass depend on the epoch of the formation, thus the lighter PBHs form earlier than the massive ones. 

For both cosmological scenarios, the following assumptions are considered throughout this work:
\begin{enumerate}
\item The distribution of the linear density fluctuations is assumed to follow Gaussian statistics. 
This has been shown to be only approximate for the actual enhanced fluctuations that are able to collapse into PBHs, and the Press-Schechter formalism can, in principle, be modified to take into account different distributions of fluctuations. Bearing this in mind but proceeding with the standard assumption, we adopt a Gaussian function $f(\nu(M))$ given by, 
\begin{equation}
f(\nu(M))=\frac{2}{\sqrt{2\pi}}\exp \left(-\frac{1}{2}  \nu(M)^{2} \right),
\label{eq: fnu Gaussian}
\end{equation}
regardless of the scale. We will further comment on this choice below.

\item In the linear regime, we associate the wavenumber of a linear fluctuation with its comoving radius, $R$, as
\begin{equation}
    k_R = \frac{2\pi}{R(M)}.
    \label{eq: k_R(M) definition}
\end{equation}

\item Given a gaussian density field, one can smooth it using a filter/window function, $\widehat{W}$. Here, we consider a top-hat $k$-space window function given by
\begin{equation}
\widehat{W}(k_R,k) = \begin{cases}
1 & \text{if $k\leq k_{R}$},\\
0 & \text{if $k> k_{R}$}.
\end{cases}
\label{eq:window}
\end{equation}

\item For both scenarios, the PBH production occurs in the radiation-dominated epoch, before the matter-radiation equality (occuring at $a=a_{\text{eq}}$); i.e, the PBH formation scale factor satisfies $a_{f}<a_{\text{eq}}$.

\item Although an extended PBH mass function covers a wide range of PBH masses, we can introduce minimum and maximum masses for the distribution; namely, $M_{\text{min}}$ and $M_{\max}$, respectively. Given that a PBH can theoretically emit particles due Hawking radiation \citep{Hawking:1974Nature,Hawking:1975}, it can be evaporated within a finite time, $t_{\text{ev}}\sim 5120 \pi\, G^2 M^{3}/\hbar c^4$. Therefore, we consider that $M_{\min}$  is the evaporation mass, $M_{\text{ev}}(z)$, for a PBH whose lifetime is the cosmic age at redshift $z$.
In principle, in the HC scenario the maximum PBH mass is given  by the horizon mass of the Universe in the matter-radiation equality epoch. In the FCT paradigm, instead, it  depends on the mechanism triggering the PBH formation.  Nevertheless, here we set the PBH maximum mass as the one which gives a unit PBH cumulative number density within the horizon volume, referred as $M_{\text{1pH}}$. Notice that both $M_{\text{ev}}$ and $M_{\text{1pH}}$ depend on the scale factor (redshift) and hence, a PBH mass function at any scale factor spans the appropriate range of masses between $M_{\text{ev}}$ and $M_{\text{1ph}}$.

\item Only a small fraction of the total energy density at early times forms BHs. This can be understood in two different ways: either all fluctuations that surpass the collapse threshold do collapse, but only the small volume corresponding to their central regions ends up forming the PBH; or only a small fraction of the fluctuations that overcome the threshold do collapse. In the first case, the fraction of the mass that ends up in the black hole is a parameter of the model, named $f_m$, and it can take different values. In the second case, $f_m = 1$ but not all densities that meet the collapse criteria form BHs. It should be mentioned that there is an implicit assumption in this prescription, namely, that there is a strong spatial correlation between the collapse of the spike-scale fluctuation and the collapse of the smaller fluctuations contained therein. In the case of $f_m<1$, this means that we assume that smaller spikes are quasi-homogeneously distributed inside the spike-scale fluctuation, and thus, when the innermost $f_m$ fraction of the Lagangian volume collapses, the same fraction $f_m$ of sub-spikes collapses as well, such that the correction is an overall factor that does not change the shape of the mass function. In the case when $f_m=1$, only a fraction of the spike-scale fluctuations does collapse, but when this happens, the entire Lagrangian volume of said fluctuations collapses. Here, we assume that in turn the same fraction  of sub-fluctuations in the entire universe collapses, such that the shape of the mass function is unaffected as well. It should be noted that global factors are immaterial as in the end the amplitude of the mass function is determined by the requirement that the total DM density should be equal to its value today.

\item For each mass function obtained from the Press-Schechter approach, we  set the normalization factor $A_n$ to ensure that the PBH mass density constitutes the present-day dark matter density.

\end{enumerate}

Finally, an additional consideration regarding the physical validity of the Press-Schechter approach needs to be satisfied. This formalism should only be used when the resulting mass function has a strong relation to the overdensities of the power spectrum of the fluctuations. 
Indeed, this will be satisfied when the overdensity for collapse is of the same order of magnitude as typical fluctuations. The latter are given by
\begin{equation}
    \left<|\delta|^2\right>^{1/2}_{a_{e}}=\left<|\delta|^2\right>^{1/2}_{a_{cmb}}\frac{a_{eq}}{a_{cmb}}\left(\frac{a_{e}}{a_{eq}}\right)^2\left(\frac{k_c}{k_{piv}}\right)^{\frac{(n_b-n_s)}{2}},
    \label{eq: typical delta FCT}
\end{equation}
where $a_{eq}$ and $a_{cmb}$ are the scale factor at equality and decoupling epochs, respectively, and $a_{e}$ is the scale factor where the typical PBH is formed. For the FCT scenario, the latter is taken as a constant, while for the HC scenario it corresponds to the average scale factor for the horizon reentry. Finally, $k_c$ is the characteristic wavenumber of the fluctuations which formed the typical PBH. For FCT it is given by the wavenumber for the PBH of average mass, whereas for HC it corresponds to the wavenumber of the horizon mass at the average formation scale factor. All averages considered in the definition of the typical fluctuation given by Eq.\eqref{eq: typical delta FCT} are mass-weighted averages, where the mass function is used as the statistical weight. Further details on this definition are discussed in \cite{Sureda:2021}.

If the overdensity for collapse is much smaller than the typical fluctuations, then simply all positive fluctuations collapse.  In this case the shape of the power spectrum has an influence on the shape of the mass function but the formalism does not inform on the condition for a positive overdensity to form a black hole.  When the overdensity for collapse is orders of magnitude above the typical fluctuations, then the collapse is stochastic and unrelated to the power spectrum. This is the case as the probability for collapse would be exponentially suppressed such that only the highest peaks in the density field could collapse, which are so scarce that the resulting mass-function is sensitive to sampling noise within the cosmic volume and also the resulting distribution would only have information of the higher-power modes of the power spectrum. 

One important point regarding the requirement that typical overdensities  be of the same order of magnitude as the collapse overdensity, is that we avoid the possible problems associated to adopting a gaussian shape for the distribution function, as the population of PBHs will be less affected by tails of the distribution.   

With these considerations at hand, we proceed to calculate the resulting mass functions.

\subsection{Construction of the mass functions for two different formation scenarios}

Under these assumptions and following the Press-Schechter approach as given by \cite{Sureda:2021} we construct an extended PBH mass function with a spike.
The resulting PBH distributions are parametrized by a \textit{characteristic mass}, $M_*$, where the distribution experiments a rapid exponential fall-off,  and the \textit{blue index}, $n_b$ which is related to the slope of the power law distribution. Our modified  primordial power spectrum with a spike imprints a significant difference with the mass function presented by \citet{Sureda:2021}.
The spike is centered on a mass, $M_{\text{p}}$, and is characterised by two parameters, its amplitude $A_2$ and width $\epsilon$. We fix the latter to $\epsilon=0.01$ for all the computations done in this work.
Given that the spectral index of the primordial power spectrum is well measured up to $k\sim1$, we  set $k_{\text{piv}}=10$.
We can use estimates of the mass function to compute the PBH number density by integrating it over a range of masses of the distribution. Additionally, the mass density of PBH in a interval of masses can be estimated by integrating $M dn/dm$. Under the scenario where PBHs  account for all the dark matter, a normalization should be calculated to ensure that the PBH mass density equals the dark matter density estimated by the cosmological observations. Further details of the normalization are described in \citet{Sureda:2021}.

We construct the spiky PBH mass functions under our two formation scenarios, horizon crossing and fixed conformal time, as follows.

\subsubsection{Horizon crossing}
A linear density fluctuation with density contrast $\delta_{f}> \delta_{c}$ will form a PBH as the fluctuation re-enters the horizon. Thus, there is a relation between the Lagrangian mass entering the horizon and the scale factor  $a_{hc}\sim M^{1/2}$. In addition, the fluctuation is related to the mass via $k_{R}=C_{hc}/\sqrt{M}$, where $C_{hc}= \pi (2H_{0}\,c\,f_m \sqrt{\Omega_{r,0}} /G)^{1/2}$. With this relation, we calculate the variance of the fluctuation and then  apply the modified Press-Schechter formalism to obtain the following HC mass function,

\begin{eqnarray}
&&\left(\frac{dn}{dM}\right)_{\text{HC}}=\frac{\rho_{\text{DM}}}{3\sqrt{2\pi}}\frac{F(M_{*})}{\, F(M)^{3}} \left[2(n_b-n_s) C_{\text{HC}}^{n_b+3} M_{\text{piv}}^{-\frac{n_b+3}{2}}+ 
\right.\nonumber\\
&&\frac{1}{2}(n_s+3)(1-n_b)C_{\text{HC}}^{n_b+3}M^{-\frac{n_b+3}{2}}+\nonumber\\
&&\frac{A_2}{A_1 \sqrt{2\pi}\epsilon}\frac{(n_s+3)(n_b+3)}{M_{\text{piv}}^{-(n_s-n_b)/2}} C_{\text{HC}}^{3+n_b-n_s} M_{\text{p}}^{-3/2}e^{\frac{9\epsilon^2}{2}}\nonumber \\ 
&& \left. \exp{\left(-\frac{\log{\sqrt{M/M_{\text{p}}}}+3\epsilon^{2}}{\sqrt{2}\epsilon}\right)^2}
 \right] e^{-\frac{1}{2}\left(\frac{\nu_{*}}{\nu_{m}}\right)^2 }
\end{eqnarray}
\noindent
where $M_{\text{p}}$ is the mass of the spike, $M_{\text{piv}}$ is the mass associated to the pivot wavenumber $k_{\text{piv}}$, and the functions $\nu_{*}$ and $\nu_{m}$ are introduced and computed in the appendix. Notice that this mass function has two terms; the first one is to the broken power law part of the spectrum, while the second one is due to the spike contribution. Finally, we notice that by setting $A_2=0$ one recovers the HC mass function found in \citet{Sureda:2021}.

\subsubsection{Fixed conformal time}
In this scenario, the wavenumber of the density fluctuations is related to the PBH mass as $k_R=C_{\text{FCT}}/\sqrt[3]{M}$, where the constant $C_{\text{FCT}}=a_{\text{FCT}}(32 \pi^4 \rho_b\, f_m/3)^{1/3}$. Here $a_{\text{FCT}}$ is the scale factor of the PBH formation which is fixed to the scale factor of the end of inflation, set to $a_\text{FCT}=10^{-26}$. Also, $\rho_b$ is the background density at the formation time, which lies well within the radiation domination epoch.
The resulting mass distribution in the FCT scenario is given by,
\begin{eqnarray}
&&\left(\frac{dn}{dM}\right)_{\text{FCT}}=\nonumber\\
&&\frac{\rho_{\text{DM}}}{3\sqrt{2\pi}}\frac{F(M_{*})}{\, F(M)^{3}} \left[(n_b+3)(n_s+3)C_{\text{FCT}}^{n_b+3} M^{-\frac{n_b+9}{3}}+
\right.\nonumber\\
&&\frac{A_2}{ A_1 \sqrt{2\pi}\epsilon}\frac{(n_s+3)(n_b+3)}{M_{\text{piv}}^{-(n_s-n_b)/3}} \frac{C_{\text{FCT}}^{3}}{M_{\text{p}}\,M^2}e^{\frac{9\epsilon^2}{2}}\nonumber \\ 
&&  \left. \exp{\left(-\frac{\log{\sqrt[3]{M/M_{\text{p}}}}+3\epsilon^{2}}{\sqrt{2}\epsilon}\right)^2}
\right] e^{-\frac{1}{2}\left(\frac{\nu_{*}}{\nu_{m}}\right)^2 },
\end{eqnarray}
where $\nu_{*}$ and $\nu_{m}$ are given in the Appendix. As in the HC scenario, when $A_2=0$, there is no  spike feature in the mass function and we recover the \citet{Sureda:2021} result.

\subsection{The maximum spike amplitude}

The spike in the mass function contains a percentage of the total mass density of the distribution.  This percentage depends on the scale of the  mass peak and its amplitude $A_{2}$. In order to fix this constant, we calculate the maximum amplitude, $A_{2,\text{max}}$, which provides the largest possible fraction of mass inside the spike, as being a function of the ratio $\rho_{\text{P}}/\rho_{\text{PBH}}$, where $\rho_{\text{P}}$ is the mass density within the spike and $\rho_{\text{PBH}}$ is the total PBH density. The latter is calculated as $\rho_{\text{PBH}}=\rho_{\text{P}}+\rho_{\text{PL}}$, where $\rho_{\text{PL}}$ is the mass density within the power law part of the mass function. We also require that $\rho_{PBH}$ does not exceed the current dark matter mass density within a $1\%$  error.

It may seem counterintuitive that one cannot simply increase $A_{2,\text{max}}$ until all of the DM is contained in the spike. However this can be understood due to the hierarchical nature of the PS mechanism. Indeed, the presence of the spike subtracts mass from the scales immediately less massive than its scale, which would have collapsed earlier, and whose fluctuations would therefore be contained within the spike. However, fluctuations with associated masses larger than the spike remain largely unaffected, as they collapse later than the spike. For this reason, the maximum fraction of DM that can be contained within the spike depends on $M_p$ and $M^*$, and it is accomplished for $M_p\simeq M^{*}$.

The result of this process allows us to understand the importance of the spike with respect to the power law part of the mass function, for different combinations of spike and characteristic masses, $M_p$ and $M_*$. 

\subsection{Analysis and examples}

\begin{figure}
  \centering
  \includegraphics[width=0.45\textwidth]{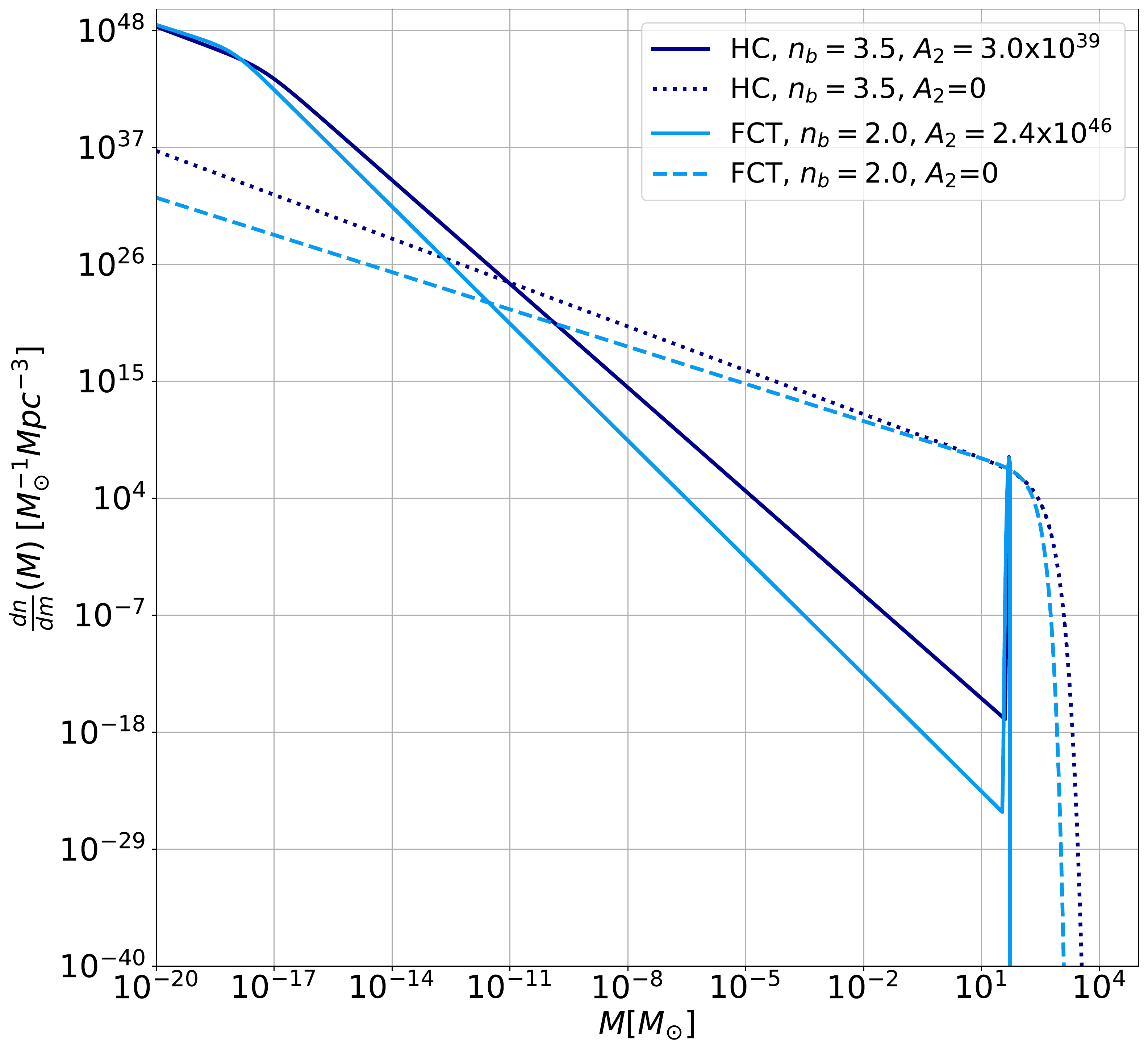}\\
   \includegraphics[width=0.45\textwidth]{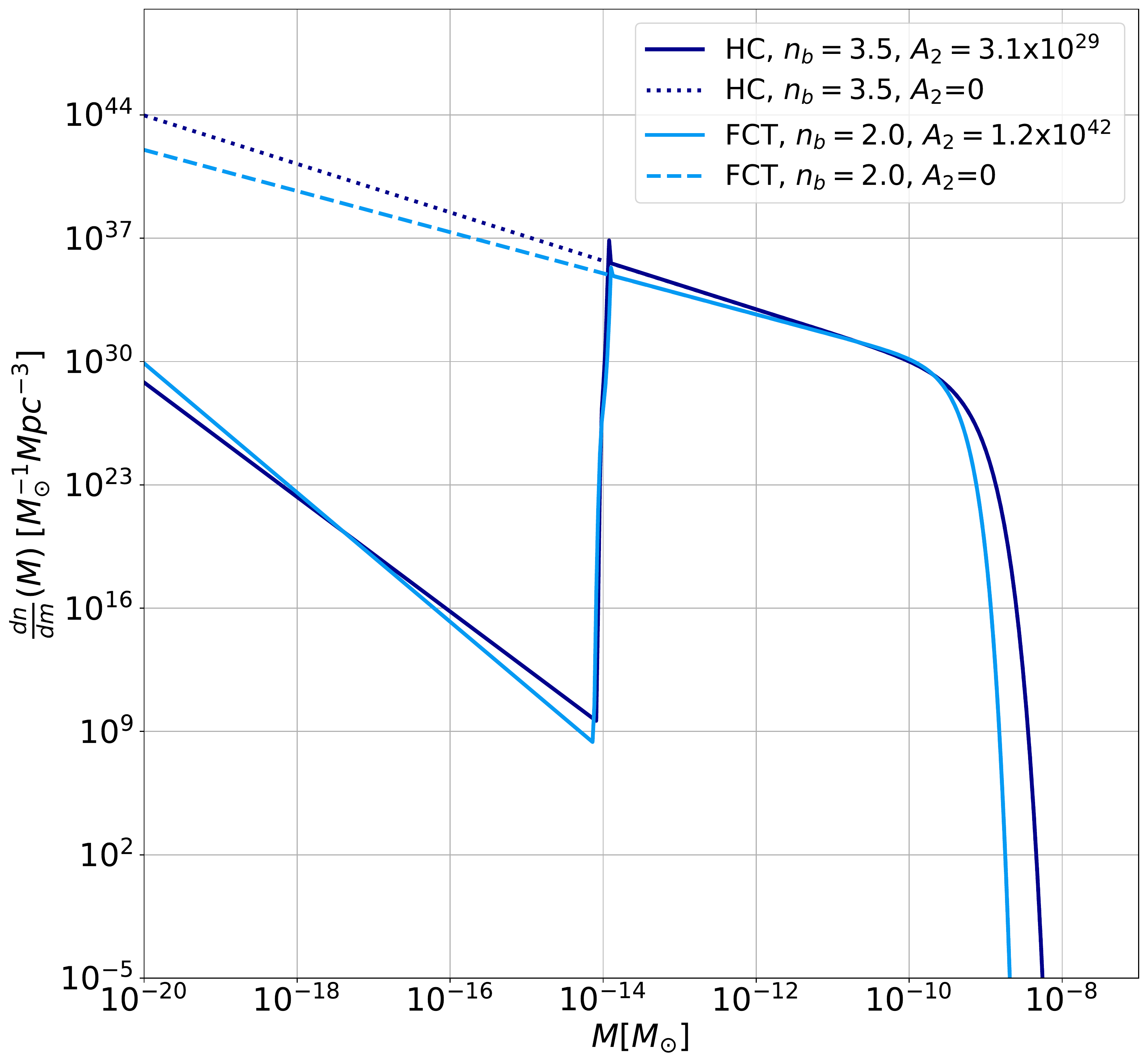}\\
  \caption{Examples of HC and FCT mass functions  for $n_b=3.5$ and $n_b=2.0$, respectively (dark and light blue lines) with exponential drop off and spike masses $M_p=M^*=50 M_{\odot}$ (top panel) and $M_p=10^{-14}M_\odot$ and $M^*=10^{-11} M_{\odot}$ (bottom panel).  We also show the extended distribution without the spike with the same $M_*$ and $n_b$ for comparison (dotted and dashed lines).}
\label{fig:dndm}
\end{figure}

\begin{table*}
\centering
\caption{Comparison of the number (columns 3, 4 and 5 and 6) and mass density (columns 7, 8, 9 and 10) of PBHs in four regions of the extended mass functions; (i) in PBHs that should have already evaporated but left Planck mass relics ($n_{relic}$ and $\rho_{relic}$), (ii) for PBHs in the log-normal spike ($n_{p}$ and $\rho_{p}$), (iii) in the power law region of the mass function in the range $M_{ev}(z=0)<M<M_{P}$ ($n_{\text{PL1}}$ and $\rho_{\text{PL1}}$), and (iv) in the power law region of the mass function in the range $M_{P}<M \le 10^{3}M_{*}$ ($n_{\text{PL2}}$ and $\rho_{\text{PL2}}$). Results correspond to a blue index $n_b=3.5$ and $n_b=2.0$ for HC and FCT, respectively, and characteristic mass scale shown in the first column.  The mass of the spike is shown in the second column. }
\resizebox{\textwidth}{!}{
\begin{tabular}{|c|c|c|c|c|c|c|c|c|c|} 
 \hline
 $M_{*}$& $M_{p}$ &$n_{\text{relic}}$ & $n_{\text{p}}$ & $n_{\text{PL1}}$&$n_{\text{PL2}}$& $\rho_{\text{relic}}$& $\rho_{\text{p}}$ &$\rho_{\text{PL1}}$&$\rho_{\text{PL2}}$ \\
$[M_{\odot}]$& [$M_{\odot}$] &[$1/\text{Mpc}^3$] & [$1/\text{Mpc}^3$] & [$1/\text{Mpc}^3$]&[$1/\text{Mpc}^3$]&  [$M_{\odot}/\text{Mpc}^3$]& [$M_{\odot}/\text{Mpc}^3$]& [$M_{\odot}/\text{Mpc}^3$]&
 [$M_{\odot}/\text{Mpc}^3$]\\ 
 \hline
\multicolumn{10}{|c|}{HC}\\
\hline
$10^{-10}$& $10^{-14}$& $1.3\times10^{31}$ & $5.7 \times 10^{22}$ & $2.2\times10^{6}$ &$1.3\times10^{22}$ & $1.5\times 10^{-7}$& $6.7\times 10^{8}$ & $1.2\times 10^{3}$& $3.3\times10^{10}$\\ 
$10^{-10}$ & $10^{-10}$& $3.9\times10^{35}$ & $1.6 \times 10^{20}$ &$2.1\times10^{28}$& $5.4\times10^{9}$ & $4.5\times 10^{-3}$& $1.6\times 10^{10}$ & $1.5\times 10^{10}$&$2.5\times 10^{-1}$\\ 
$50$& $25$& $3.5\times10^{25}$ & $6.2 \times 10^{8}$ & $5.4\times10^{16}$&$2.6 \times 10^{8}$&$7.2 \times 10^{-15}$& $1.8 \times 10^{10}$& $1.2 \times 10^5$& $1.6\times10^{10}$\\ 
$50$& $50$& $3.1\times10^{35}$ & $3.2 \times 10^{8}$ & $1.8\times10^{28}$ &$4.0 \times 10^{-17}$&$3.5 \times 10^{-3}$& $1.6 \times 10^{10}$&$1.6\times 10^{10}$& $2.0\times10^{-15}$\\ 
 $100$ & $50$& $4.2\times 10^{23}$& $3.1\times 10^{8}$&$3.5\times10^{16}$ & $1.3\times10^{8}$&$4.7\times10^{-15}$& $1.8\times10^{10}$& $1.2\times10^{5}$&$1.6\times10^{10}$\\
 $100$ & $100$& $3.2\times 10^{35}$& $1.6\times 10^{8}$&$1.9\times10^{28}$ &$7.9\times10^{-18}$&$3.6\times10^{-3}$& $1.6\times10^{10}$& $1.6\times10^{10}$&$8.1\times10^{-16}$\\
 \hline
 \multicolumn{10}{|c|}{FCT}\\
\hline
$10^{-10}$& $10^{-14}$& $6.0\times10^{25}$ & $4.8 \times 10^{22}$ & $4.6\times10^{6}$&$4.6\times10^{21}$ & $6.6\times 10^{-13}$& $6.1\times 10^{8}$ & $8.7\times 10^{1}$& $3.3\times 10^{10}$\\
$10^{-10}$& $10^{-10}$& $1.9\times10^{32}$ & $1.6 \times 10^{20}$ & $3.8\times10^{28}$ & $8.6\times 10^{5}$& $2.1\times 10^{-6}$ & $1.6\times 10^{10}$& $1.5\times10^{10}$&$8.8\times10^{-5}$\\
$50$& $25$& $1.0\times10^{16}$ & $5.4 \times 10^{8}$ & $7.5\times10^{12}$ & $2.8\times 10^{8}$& $1.2\times 10^{-22}$ & $1.8\times 10^{10}$&$6\times10^{3}$&$1.6\times10^{10}$\\
$50$& $50$& $1.8\times10^{32}$ & $3.2 \times 10^{8}$ & $3.7\times10^{28}$ & $6.2\times 10^{-26}$& $2.0\times 10^{-6}$ & $1.6\times 10^{10}$&$1.5\times10^{10}$&$3.2\times10^{-24}$\\
 $100$ & $50$& $6.0\times 10^{15}$& $2.8\times 10^{8}$&$4.2\times10^{12}$& $1.4\times 10^{8}$& $6.7\times 10^{-23}$ & $1.8\times 10^{10}$& $2.4\times10^{5}$&$1.6\times10^{10}$\\
 $100$ & $100$& $2.0\times 10^{32}$& $1.6\times 10^{8}$&$3.9\times10^{28}$ & $8.1\times 10^{-25}$& $2.2\times 10^{-6}$ & $1.6\times 10^{10}$ &$1.5\times 10^{10}$&$8.3\times10^{-25}$\\
 \hline
\end{tabular}}\label{tab:relics}
\end{table*}

\begin{figure*}
  \centering
  \includegraphics[width=0.33\textwidth]{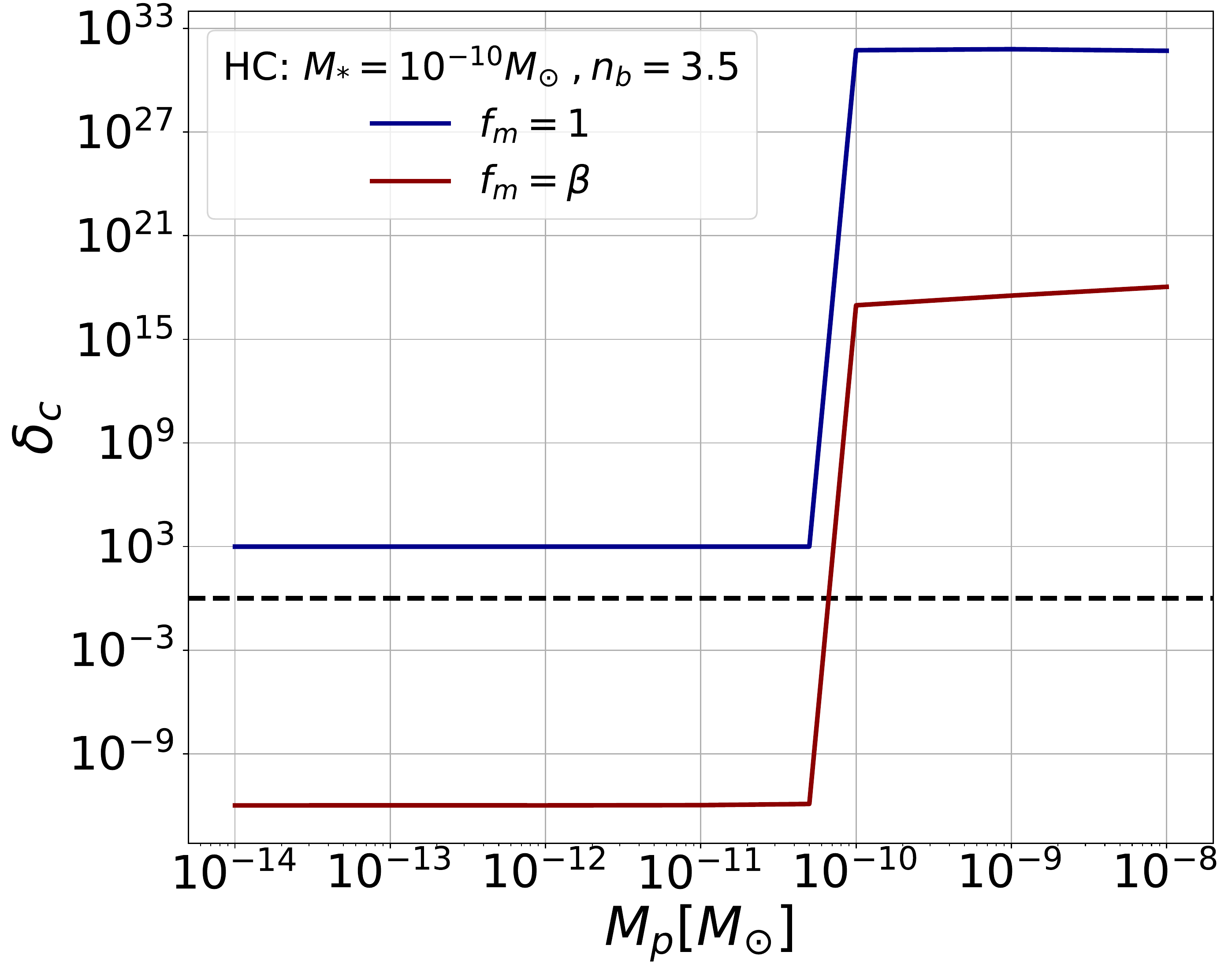}
    \includegraphics[width=0.33\textwidth]{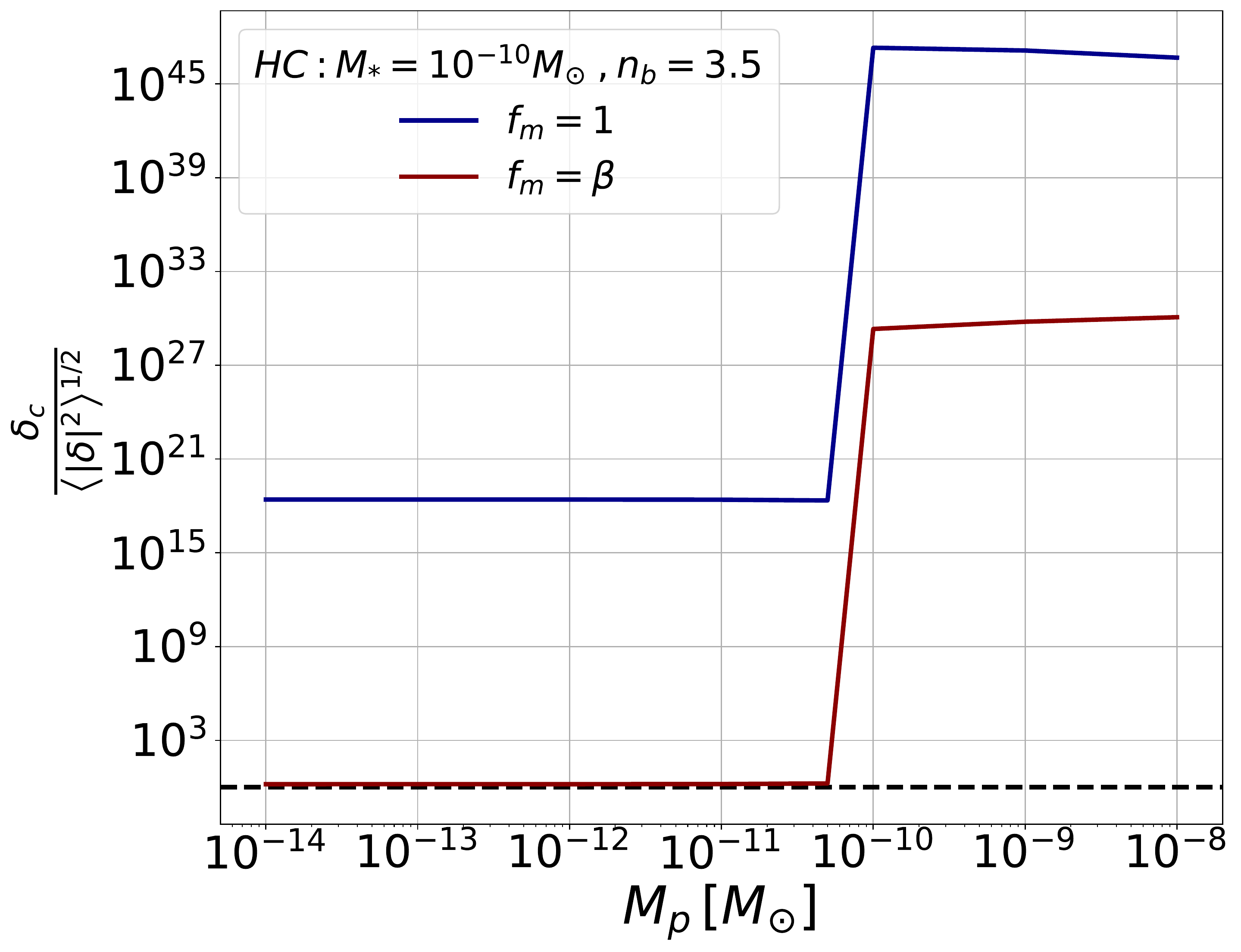}
  \includegraphics[width=0.33\textwidth]{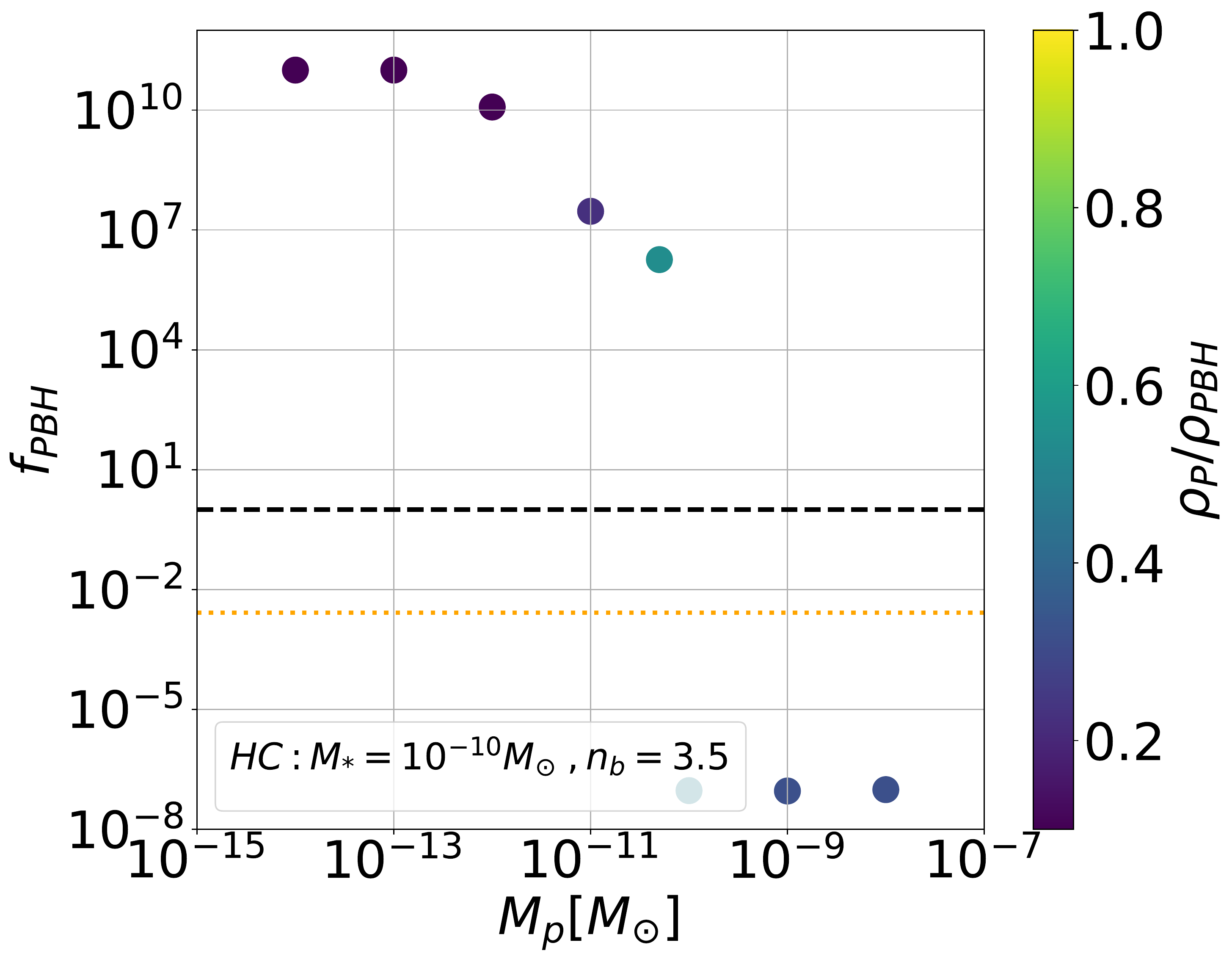}\\
   \includegraphics[width=0.33\textwidth]{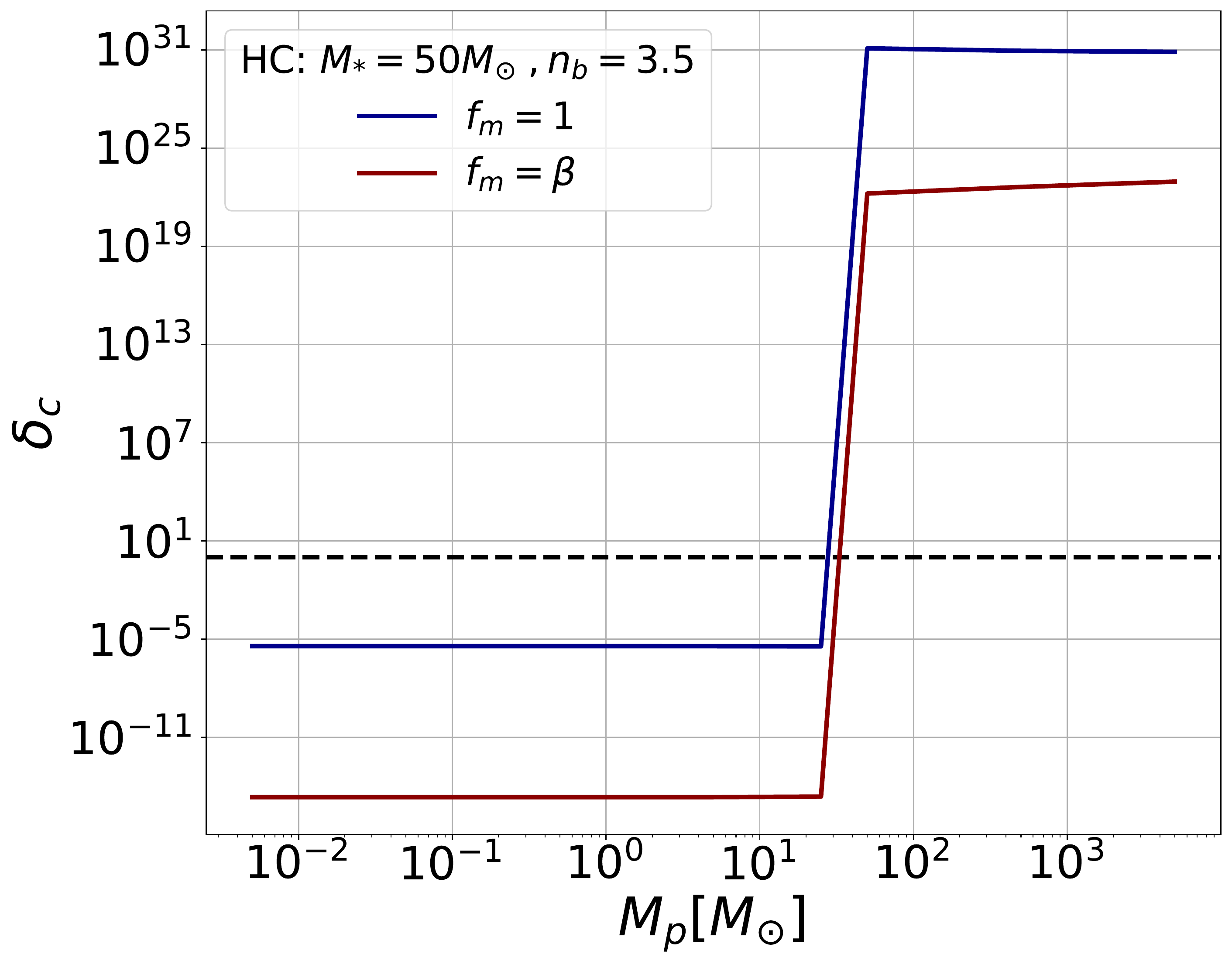}
   \includegraphics[width=0.33\textwidth]{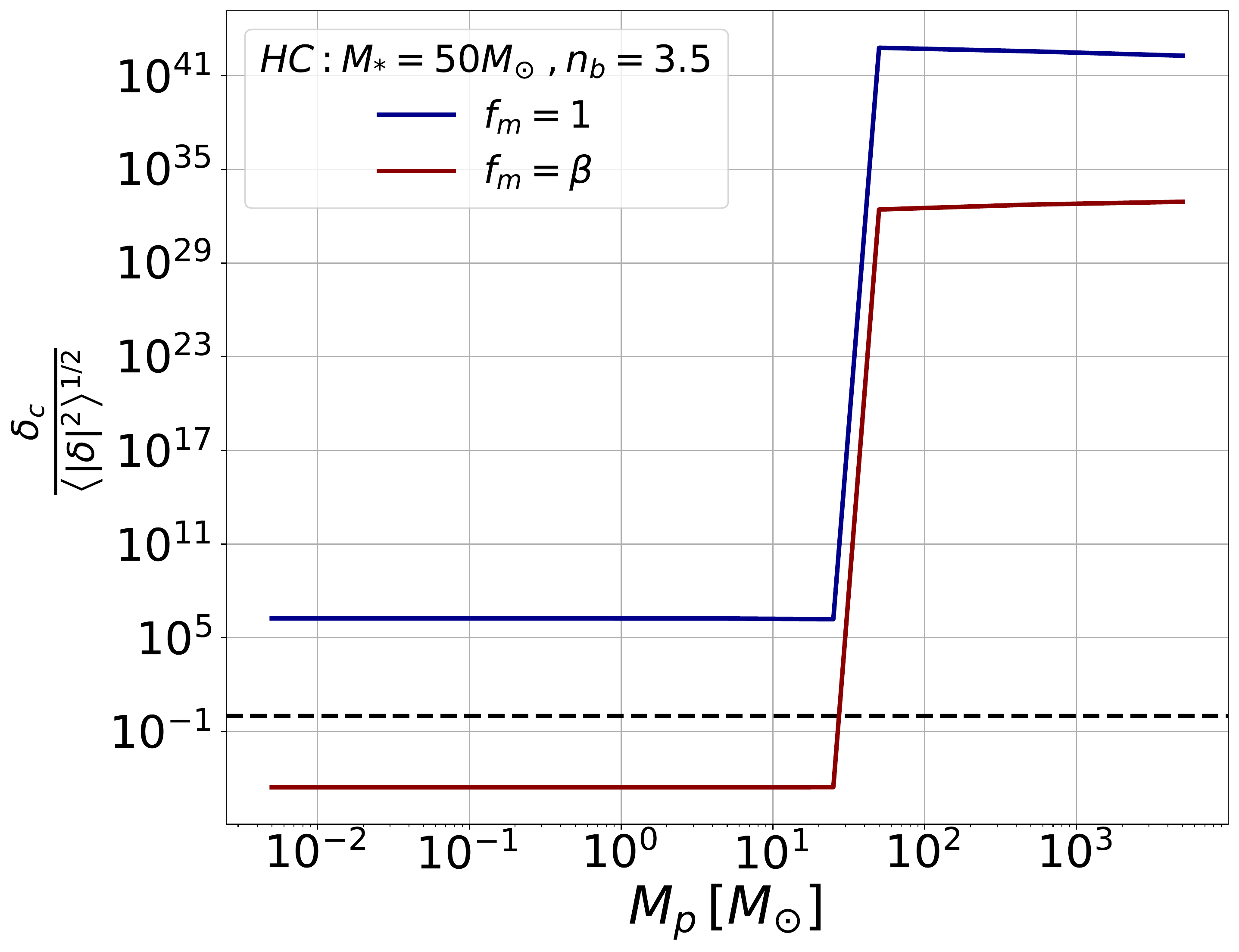}
   \includegraphics[width=0.33\textwidth]{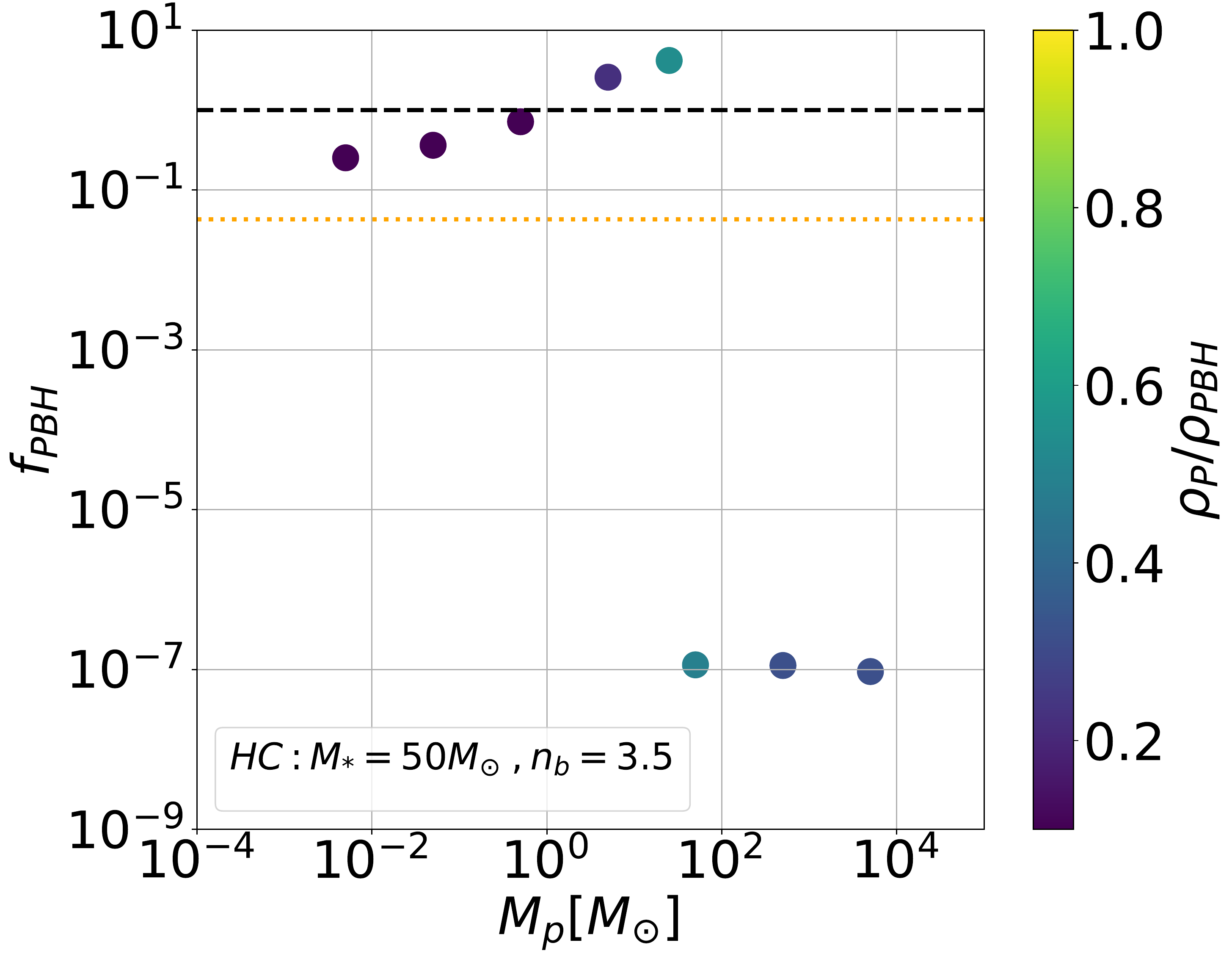}\\
   \includegraphics[width=0.33\textwidth]{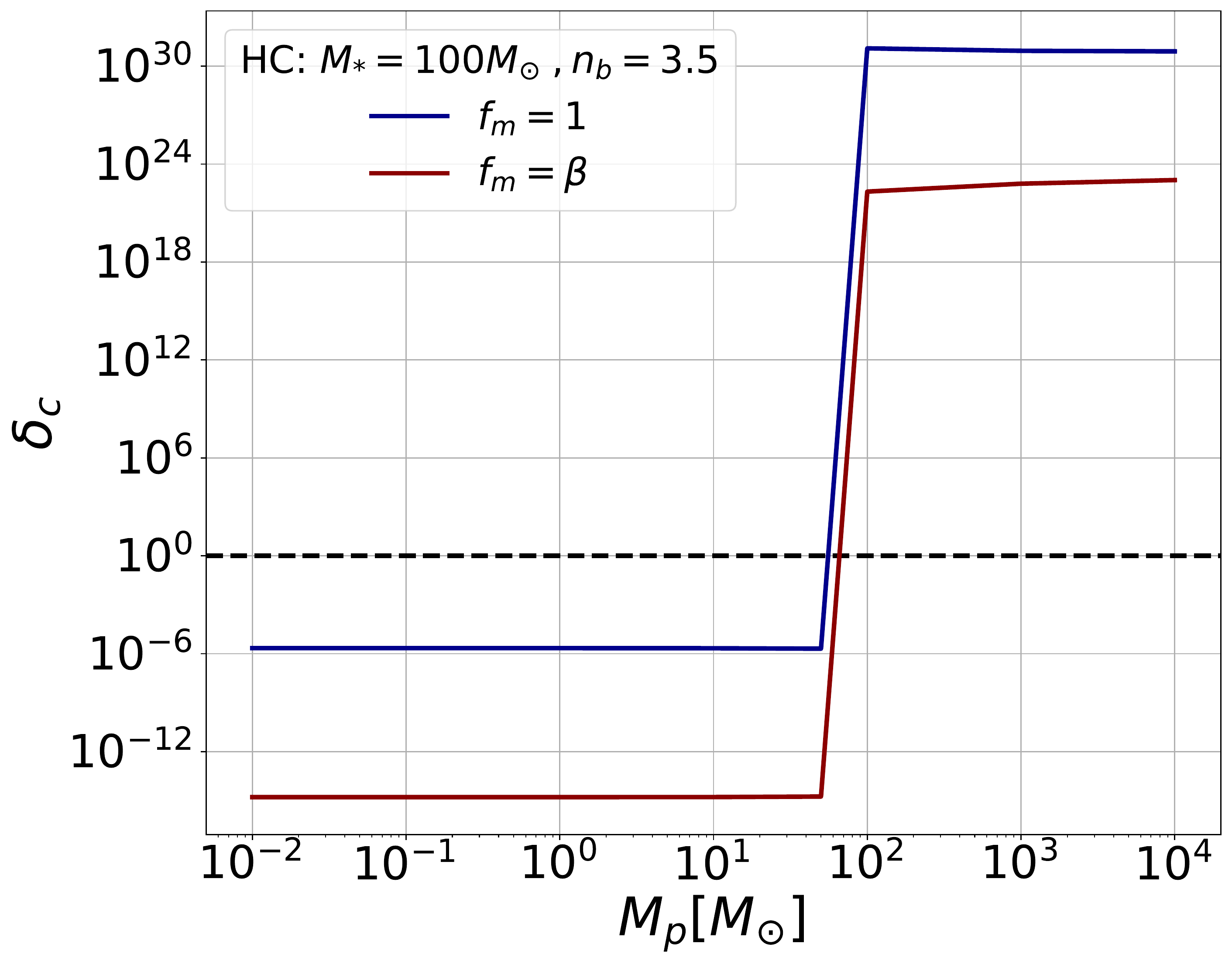}
    \includegraphics[width=0.33\textwidth]{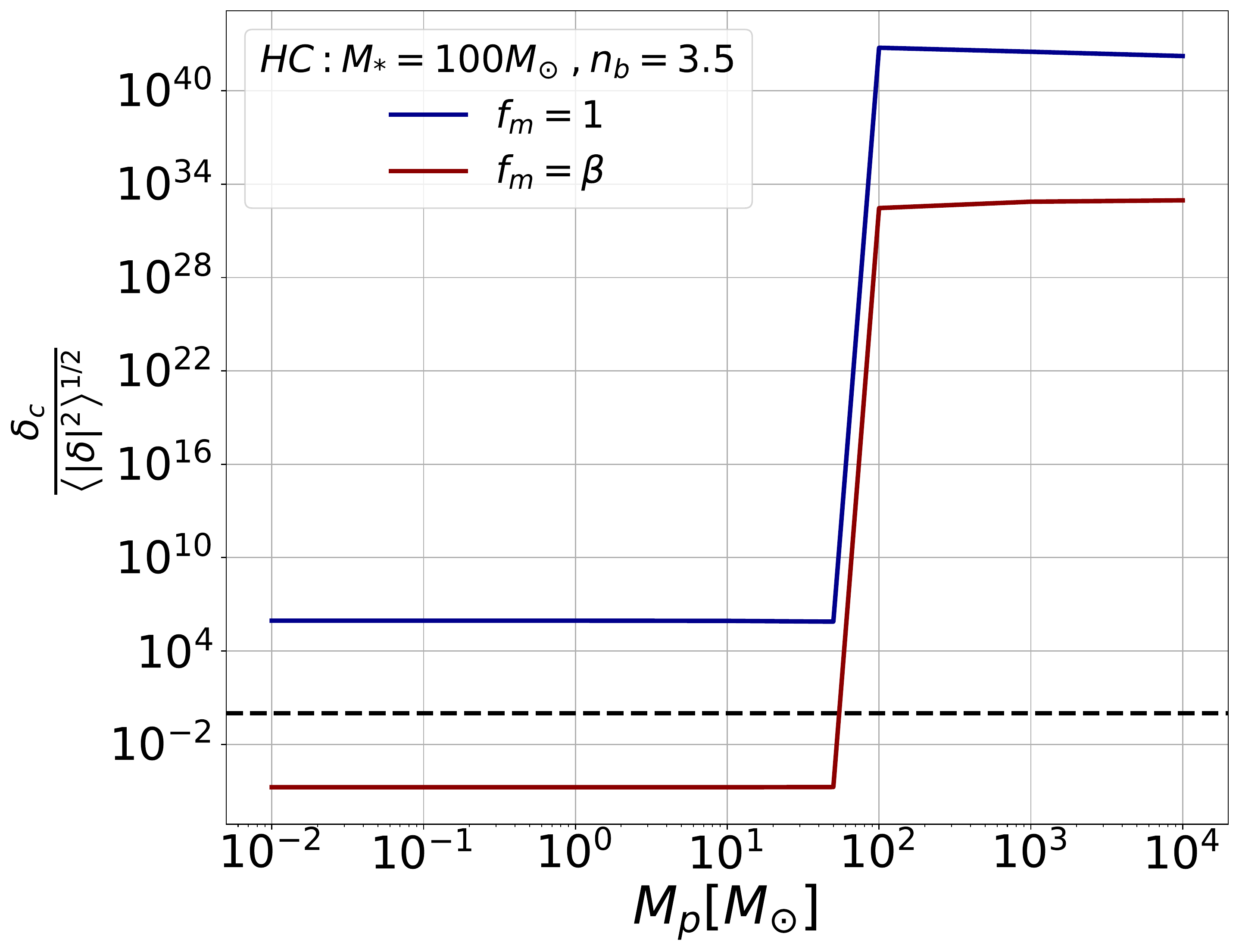}
   \includegraphics[width=0.33\textwidth]{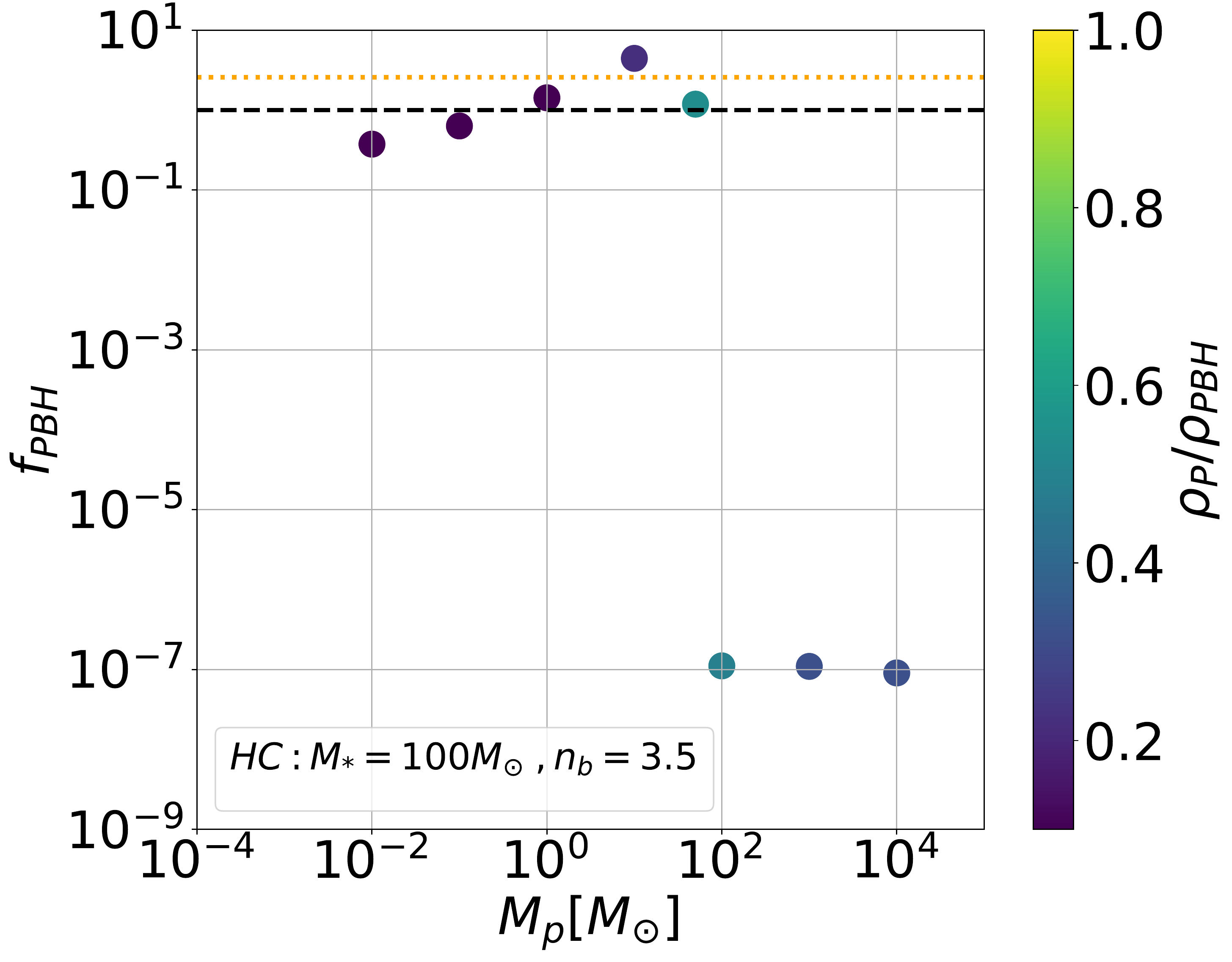}\\
\caption{Critical overdensity for collapse (left column), ratio of the threshold density over the amplitude of the typical density contrast (centre column), and fraction of DM in PBHs (right column) for extended mass distributions with a spike as a function of the peak mass $M_{\text{p}}$, in the HC scenario.  Top, middle and bottom rows correspond to different values of characteristic mass $M^*$ as indicated in the legend.  Different line colours in left and centre columns indicate different choices of $f_{m}$.  On the right column, the colour of the dots corresponds to the fraction of mass density in PBHs within the spike, as indicated by the colour bar on the right, for $f_m=1$.  On this column, the yellow dashed horizontal line shows the fraction of DM in PBHs for the extended mass function without the spike.  In all panels the solid horizontal line shows the unit value.} 
\label{fig:HC}
\end{figure*}

\begin{figure*}
  \centering
  \includegraphics[width=0.33\textwidth]{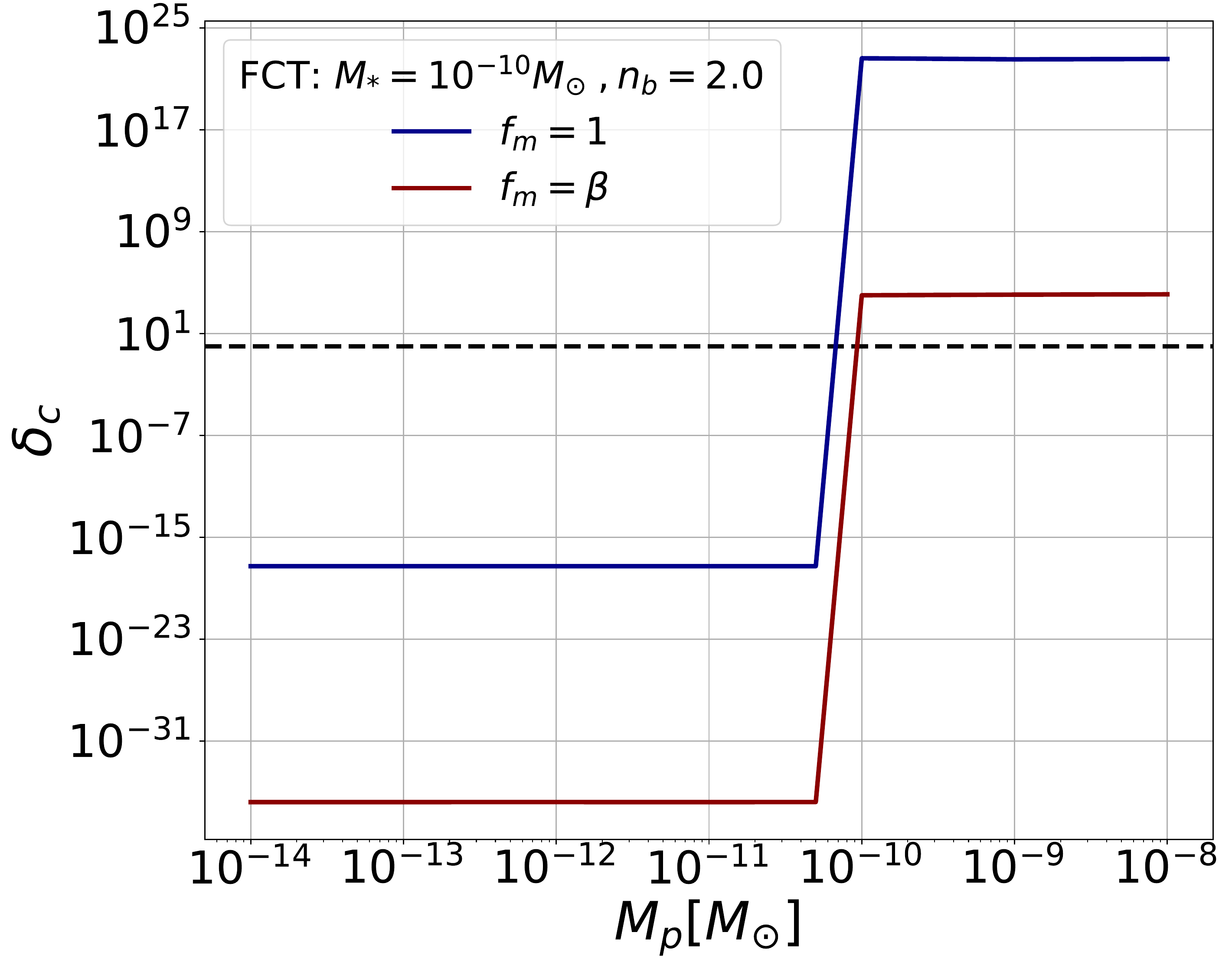}
    \includegraphics[width=0.33\textwidth]{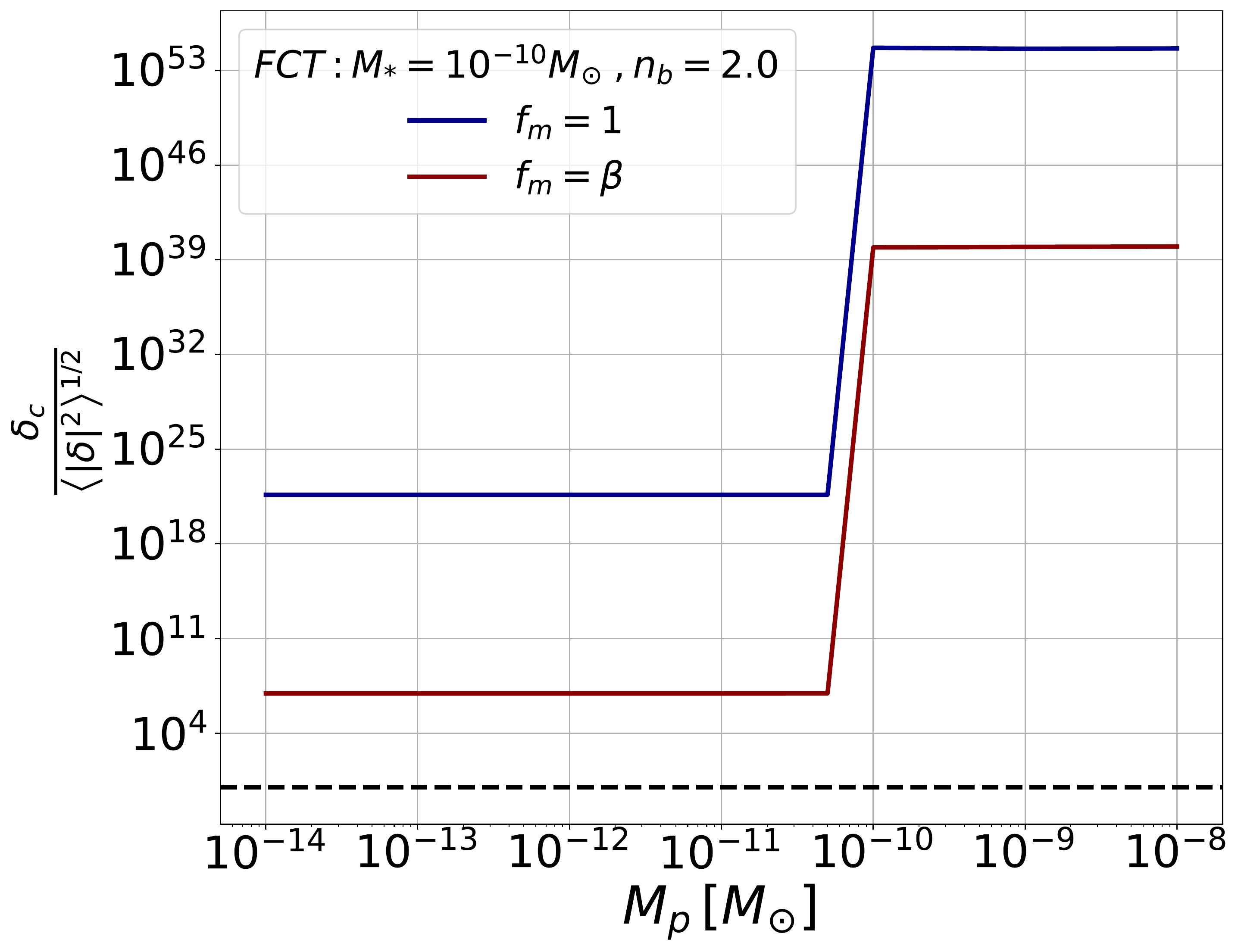}
  \includegraphics[width=0.33\textwidth]{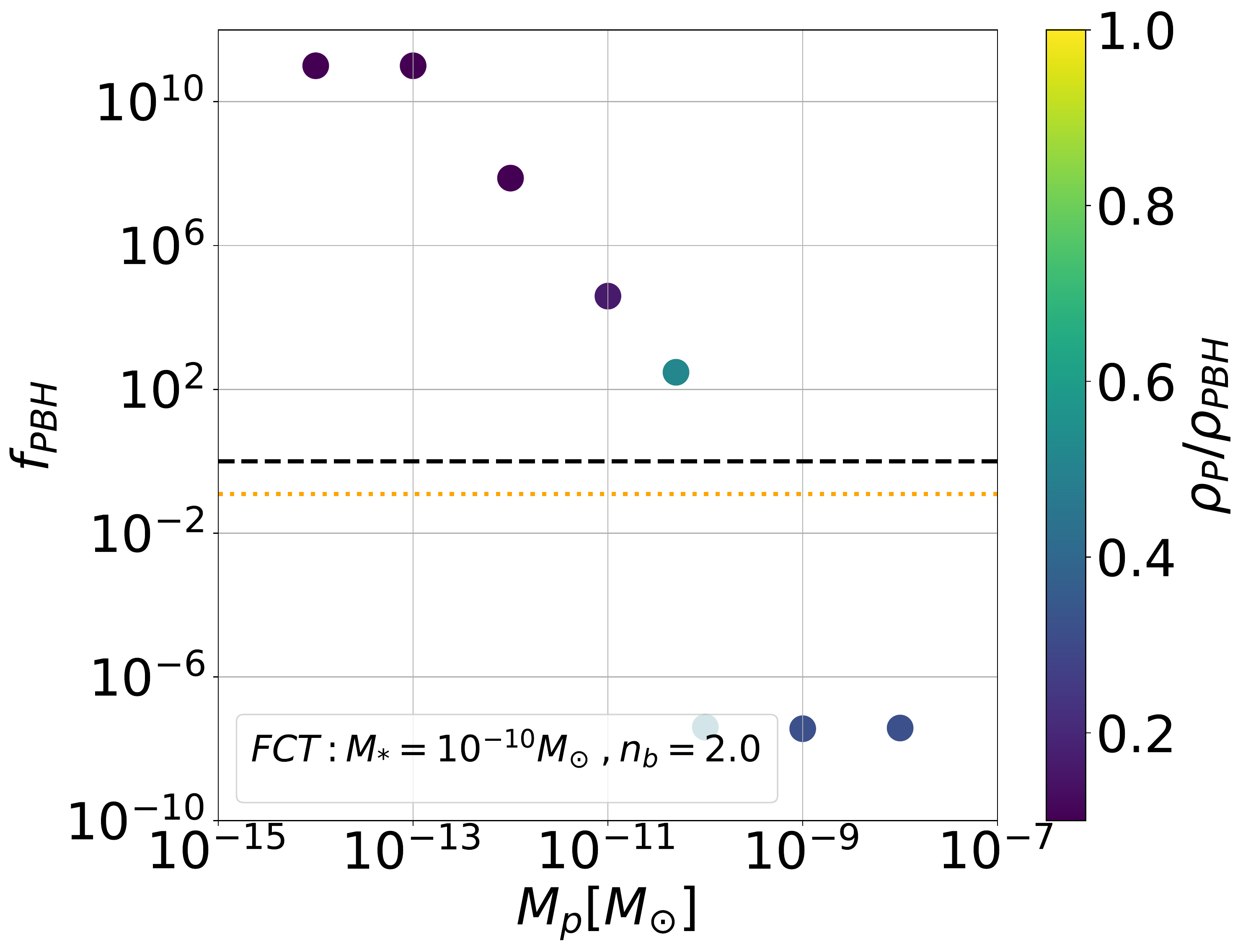}\\
   \includegraphics[width=0.33\textwidth]{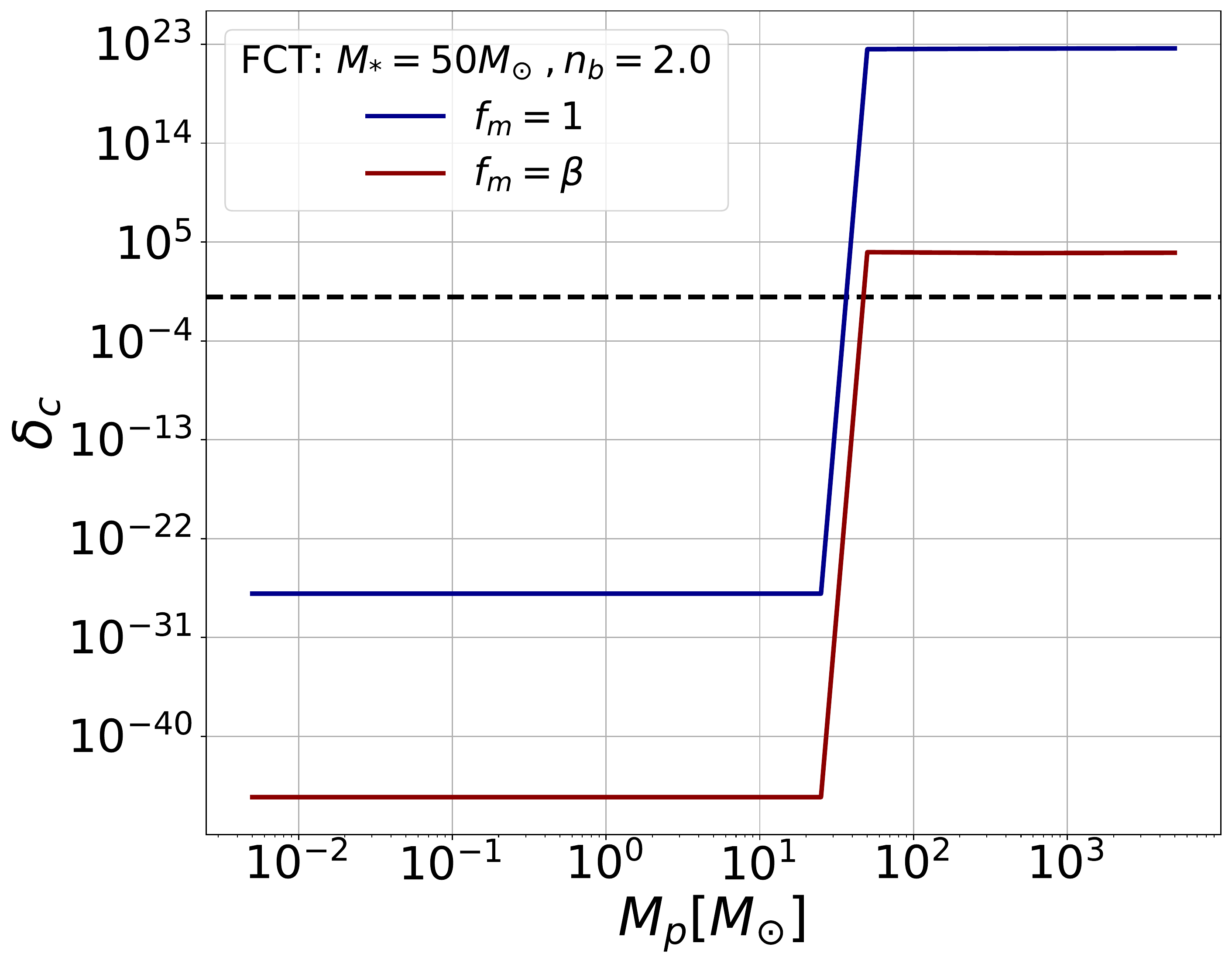}
   \includegraphics[width=0.33\textwidth]{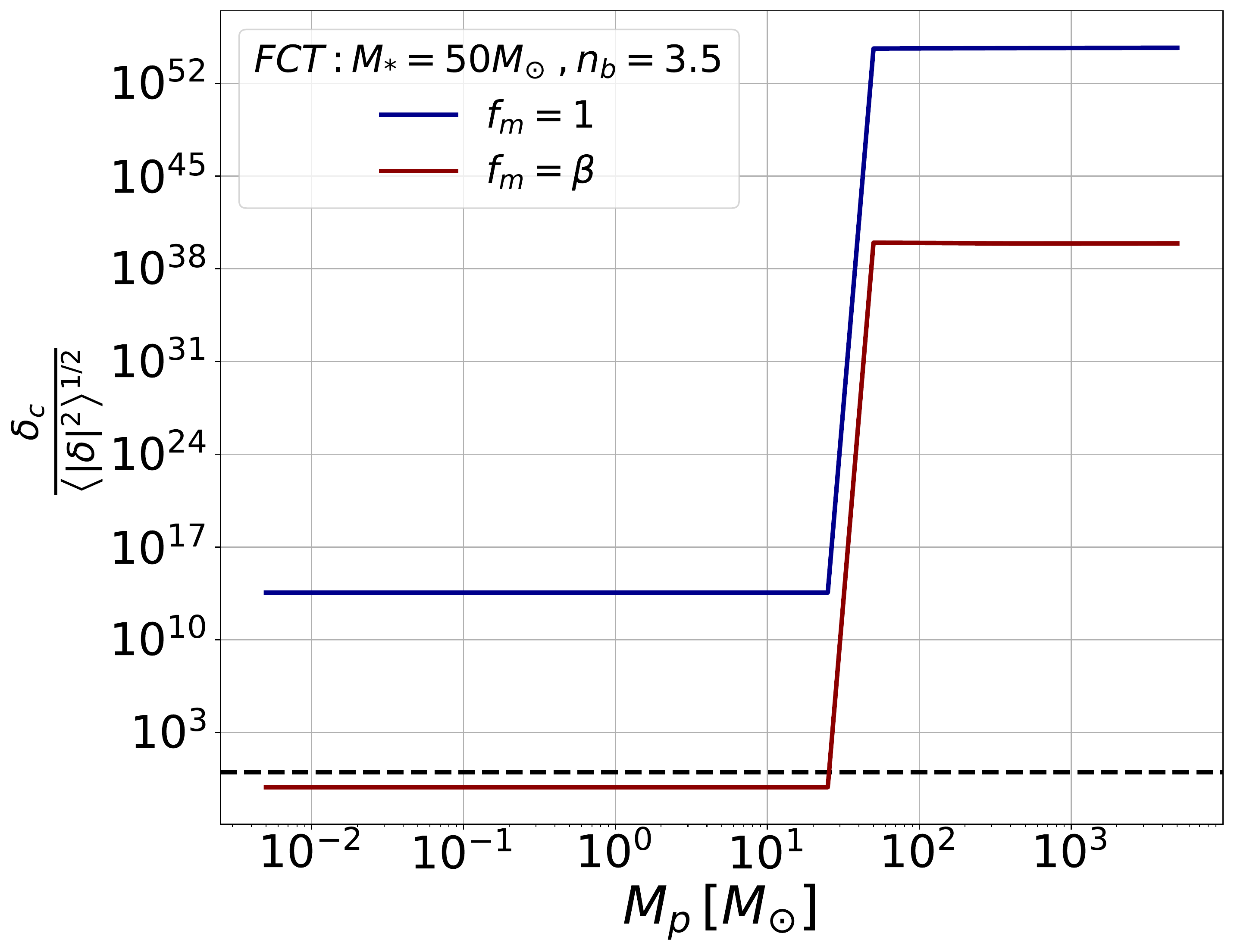}
   \includegraphics[width=0.33\textwidth]{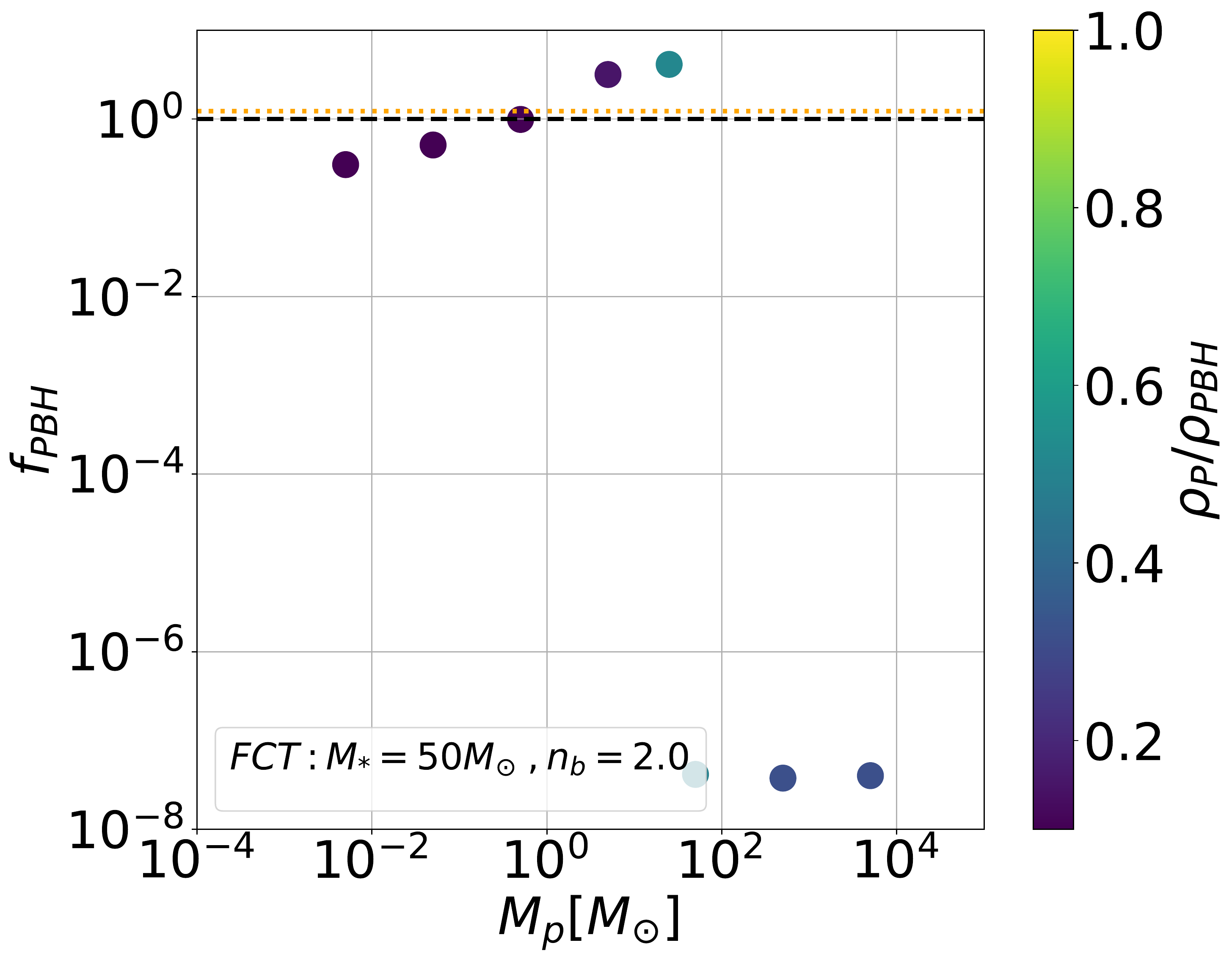}\\
   \includegraphics[width=0.33\textwidth]{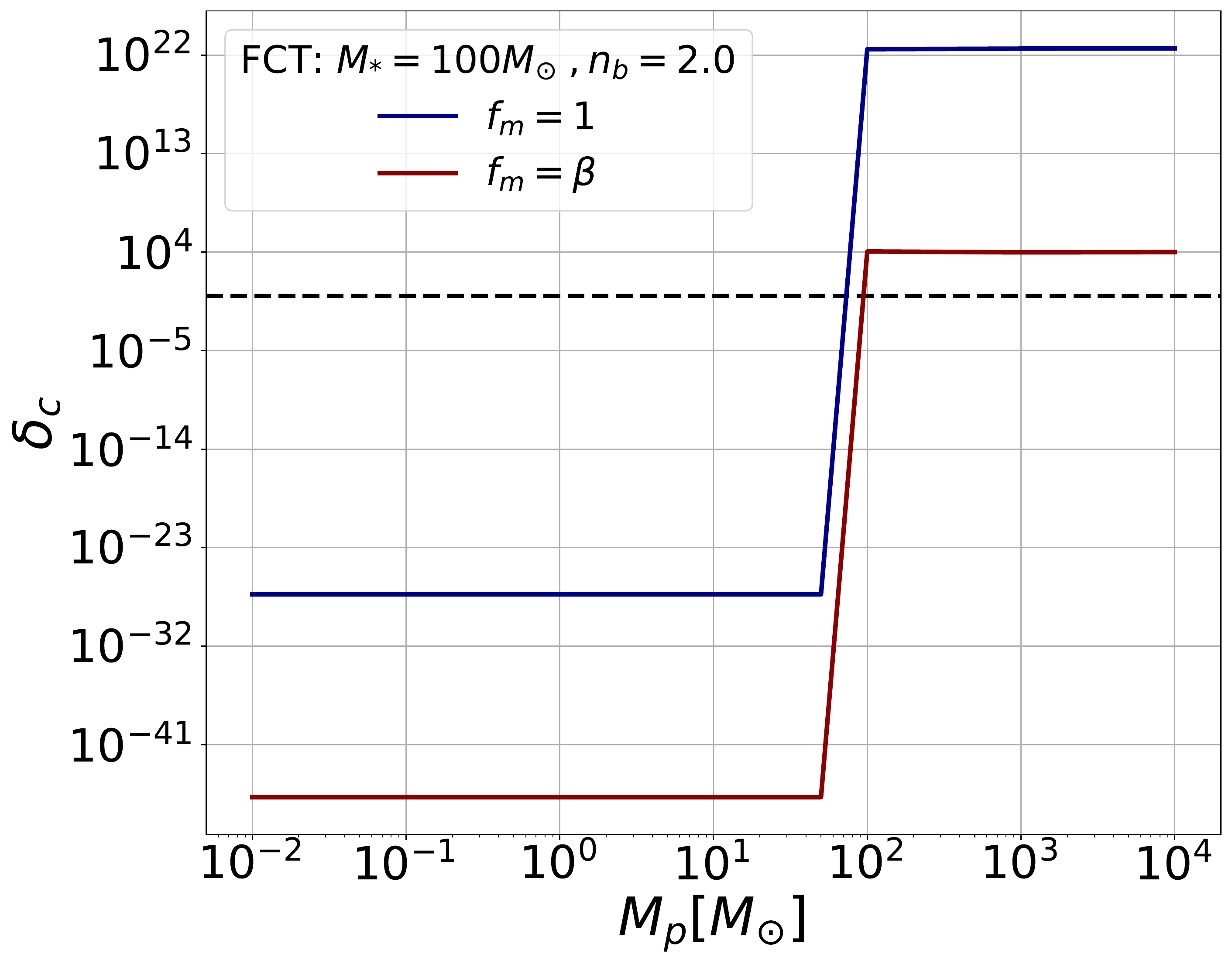}
    \includegraphics[width=0.33\textwidth]{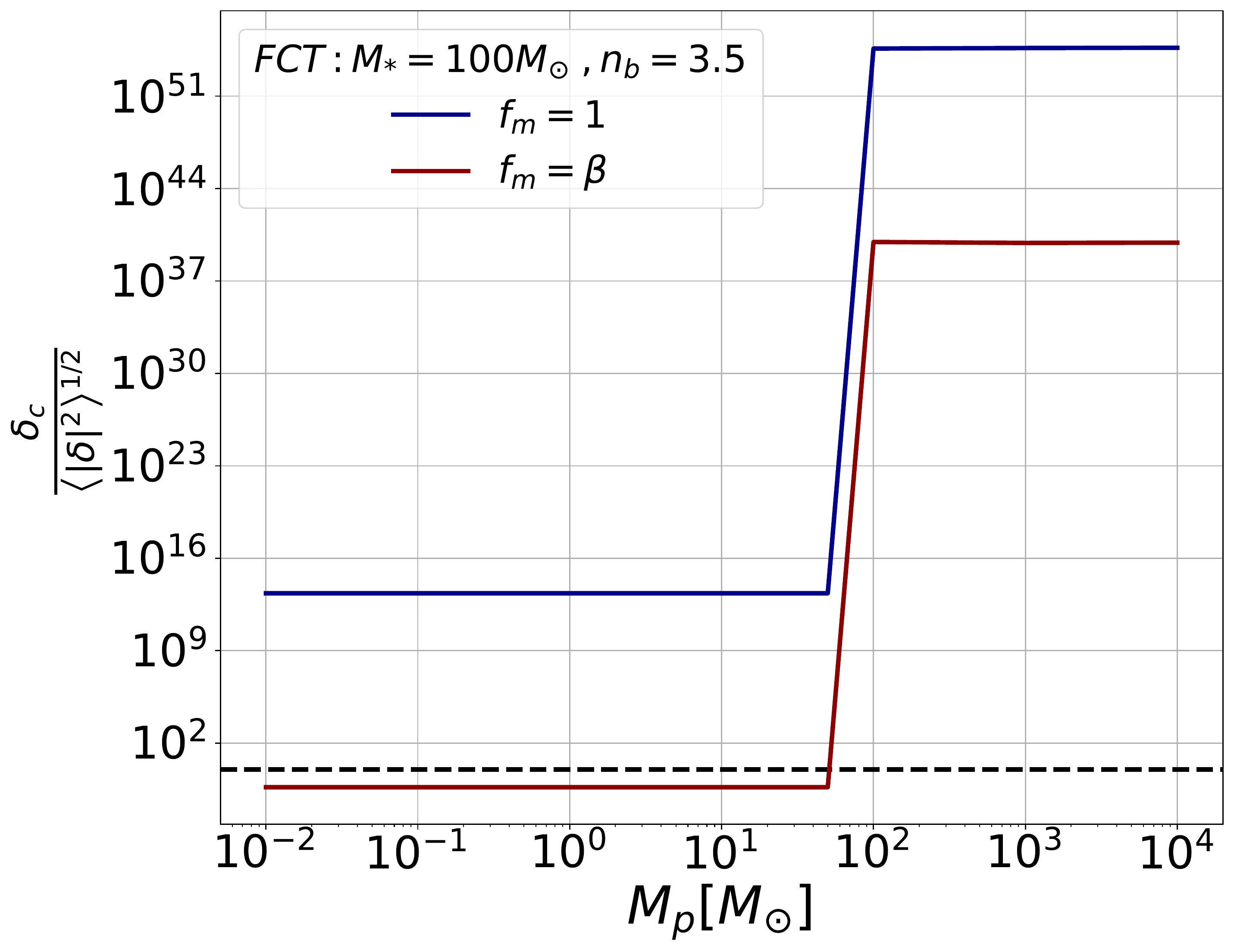}
   \includegraphics[width=0.33\textwidth]{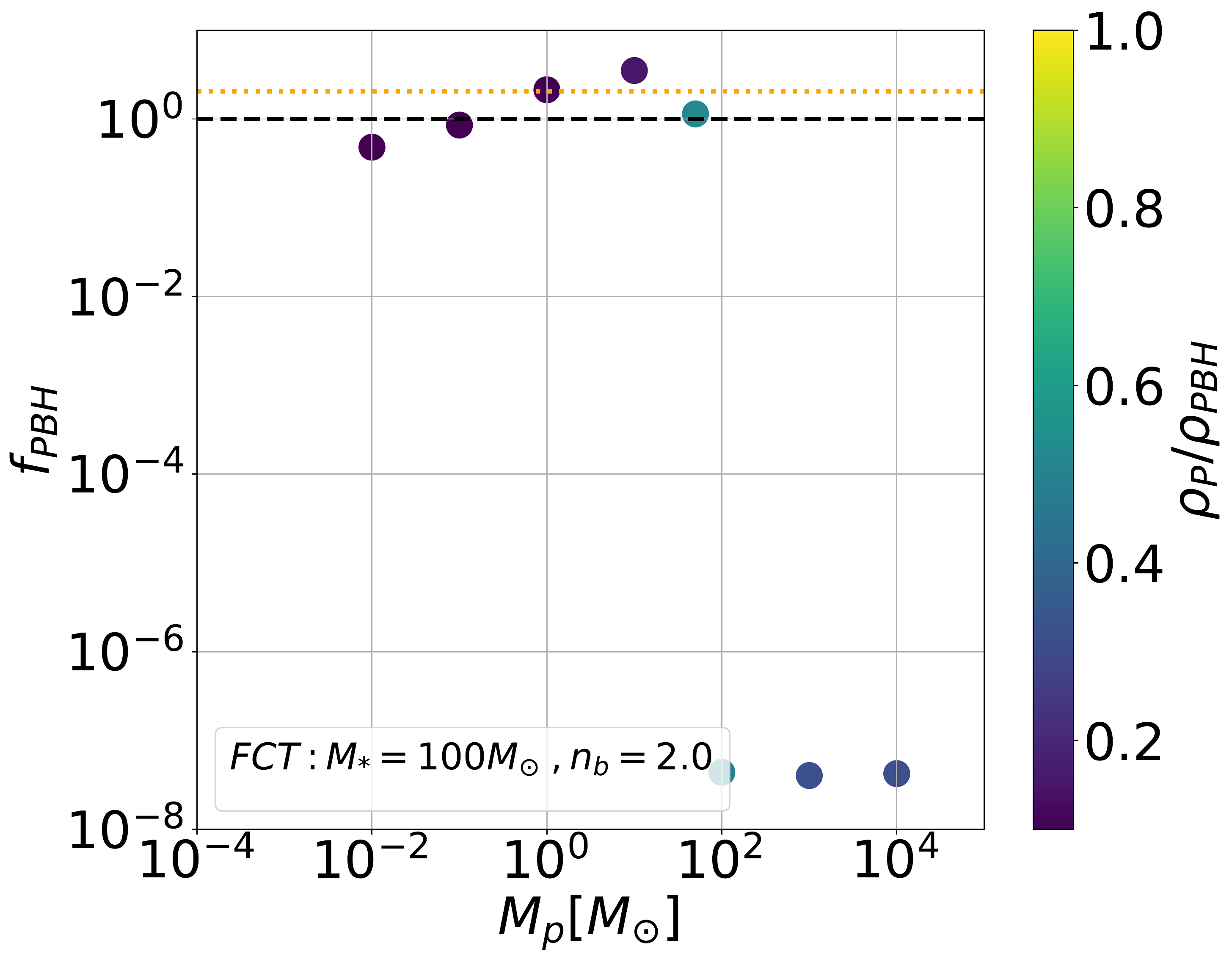}\\
\caption{Same as Fig. \ref{fig:HC} for the FCT scenario, with $n_b=2.0$.}
\label{fig:FCT}
\end{figure*}

We show examples of the resulting mass functions in Fig. \ref{fig:dndm}, for both the HC and FCT cases, with the spike that contains the maximum fraction of mass density (solid) and without a spike (dotted and dashed lines). As it can be seen, the presence of the spike tends to make the abundance of lower PBH masses to significantly decrease when compared to the case without a spike. For instance, even at the mass of PBHs that are evaporating today ($\sim 10^{-20}M_\odot$), the amplitude of the mass function with a spike can be about $15$ orders of magnitude lower in the HC case with $M_*=10^{-11}M_\odot$ (see bottom panel of Fig. \ref{fig:dndm}). Since BH formation at horizon crossing is a hierarchical process, once a region collapses, all fluctuations inside its volume collapse as well and are incorporated into the larger fluctuation that contains them. Therefore, a spike at a certain scale will partially deplete fluctuations at smaller scales as the excess density at the larger scale will incorporate smaller fluctuations that are removed from the statistical sample.
Although the fluctuations at scales that are just smaller than the spike are depleted, because the total DM density is held fixed, conservation of mass imposes an increase in the abundance at the smallest scales, which is reflected in a change of the mass-function slope.

When $M_p\ll M_*$, the spike mass almost acts effectively as a lower limit for the PBH mass, as PBHs with lower masses than $M_p$ show a dramatic drop in abundance; i.e. the mass function resembles a power law with lower and upper limits in PBH mass.  
The logarithmic slopes  at masses lower than the characteristic mass $M_*$ depend on the blue index as $M^{-(nb+3)/2}$ and $M^{-(nb+9)/3}$ for HC and FCT scenarios respectively. These slopes come from the combinations of the exponents of the factor $F(M)$ (see Eq.\eqref{eq:gM} in the Appendix) and those in the other terms of the mass function. When $A_2$ is different of zero, the function $F(M)$ behaves as a constant whereas when $A_2$ equals zero, it behaves as a power law and the slopes are consistent with those found in the previous work by \citet{Sureda:2021} i.e. $M^{-(9-nb)/4}$ and $M^{-(9-nb)/6}$ for HC and FCT respectively.

The first entries in the HC and FCT sections of Table \ref{tab:relics} show the number and mass density in different regions of  mass functions with $M_p<<M_*$\footnote{Notice that only number densities above $\sim 10^{-12}$Mpc$^{-3}$ ensure that such PBHs can be found within the present-day Hubble Volume.  This is the case of almost all the examples shown in the table, except for some of the abundances of PBHs with masses larger than $M_p$.}.  In this table we separate the mass function in four parts. (i) As it is not completely clear whether PBHs with masses below the present-day evaporation mass have evaporated completely or if they could have left a relic behind (See for instance \citealt{Araya:2022prep}) the first part of the mass function we explore corresponds to PBHs that should have evaporated by the present day; here we calculate the total number density of black holes below the evaporation mass; the matter density for the relics is calculated assuming that their mass is the Planck mass. (ii) The spike, i.e. all PBHs within the log-normal spike of the mass function, (iii) the extended part consisting of the power law from the evaporated mass up to the spike mass, and (iv) the power law starting at the spike mass up to the exponential cut-off.   As can be seen, for the case with $M_p<<M_*$ the spike contains similar number density (although much lower mass density) as the region with higher masses than the spike, leaving a small number of PBHs with lower masses.  The spike effectively concentrates the number and mass density of the mass function of PBHs less massive than $M_p$.  The only exception are relics which would be more numerous than the spike or the extended part, but that would contain the smallest fraction of mass density.

On the other hand, when $M_p\simeq M_*$, there seems to be little difference between a monochromatic distribution and the extended$+$spike ones (cf. top panel of Figure \ref{fig:dndm}); also in this case, the number density of the power-law and exponential drop off of the mass function contains a negligible mass density in PBHs.  
However, depending on the exact relative values of $M_p$ and $M_*$, the population of the extended distribution can be as important as that of the spike, particularly for PBH with masses above that of the spike.  This can be clearly seen in the remaining entries of Table \ref{tab:relics}, which correspond to different values of $M_*$ with $M_p\sim M_*$, for HC and FCT. 
{Even though in this case, by construction, more than $50\%$ of the mass density is in the mass function spike, the number of relics can be of up to $15$ orders of magnitude larger than the number of PBHs in the spike (similar to the case of spike mass much lower than characteristic mass).  The number density of PBHs in the power law region of the mass function can either be negligible,  similar, or several orders of magnitude more numerous than PBHs in the spike, depending on how similar are the values of $M_p$ and $M_*$.  In particular, the population of non-evaporated PBHs in the power law region more massive than $M_p$ remains important, both in number and mass, compared to the spike when $M_p<M_*$.  
If relics do indeed exist, the mass density locked with these is several orders of magnitude lower than in the power law regions or in the spike, regardless of the mass function parameters.}  

The prominence of different populations of PBHs in addition to the one corresponding to the monochromatic spike, are predictions for the specific extended functions with spikes presented here, and show that there could  be additional significant populations of PBHs of larger mass than the spike, and of small PBHs and even relics, if the latter do indeed exist.

Note that the mass distributions with spike shown in this figure were chosen since their threshold density is $\delta_{c}\sim 1$, actually similar to the value inferred from numerical simulations \citep[see][and references therein]{Nadezhin:1978SvA,Novikov:1980SvA,Niemeyer:1998PhRvL,Niemeyer:PRD99,Harada:2013PhRvD,Musco:2019,Young:2019JCAP,Escriva:2020,Musco:2021}. In addition, the ratio of this density over the typical density of the fluctuations is also $~1$ when $M_p\sim M_*$, for which the conditions for a statistically meaningful use of Press-Schechter are satisfied.  

\section{PBHs from extended mass distributions with a spike as dark matter}\label{sec:constraints}

There are several ways to use current observations to constrain monochromatic PBH mass functions as dark matter. For instance, \cite{Hawking:1975} postulated that a BH could radiate particles with a temperature $T_{H}\sim M^{-1}_{BH}$. If a PBH evaporates under this hypothetical Hawking radiation mechanism, the emitted radiation could affect the synthesis of light elements in the primeval Universe \citep{Carr:2020_constraints,Arbey:2020PhRvD}. The energy emission by evaporating PBHs can source the gamma ray emission from the galactic center and diffuse extragalactic gamma ray background \citep{Carr:2016_gammaray,Arbey:2020PhRvD,Chen:2022PhRvD,Mosbech:arXiv220305743M,Laha:2019PhRvL.123y1101L,Laha:2020PhRvD.101l3514L}. Thus, the PBH evaporation by Hawking radiation imposes constraints on lighter PBHs. 

On the other hand, gravitational lensing effects by PBHs of different light sources and their dynamical effects on galactic disks and binary systems put limits on PBH with intermediate masses \citep{Carr:1999ApJ_dynamical,Tisserand:2007,Green:2016PhRvD}. 

Recently, \citet{Sureda:2021} proposed that the mass function of supermassive black holes in the center of galaxies can be used to provide constraints on large mass PBHs.
Also, the abundance of halos and their clustering can bound extended mass distributions with large PBH characteristic mass \citep{Padilla:2021}. \citet{Carr:2020_slargepbhs} provide further constraints on PBHs in the mass range $10^{12}-10^{18} M_{\odot}$ using dynamical, accretion and emission effects.

All these observables can constrain 
the fraction of mass density of PBHs as dark matter given by
\begin{equation}
    f_{\text{PBH}} = \frac{\rho_{\text{PBH}}}{\rho_{\text{DM}}},
    \label{eq:fraction_of_DM_in_PBH}
\end{equation}
as a function of the (monochromatic) PBH mass in the range $10^{-20}-10^{15} M_{\odot}$.
Notice that $f_{\text{PBH}}$ is different from $f_m$. The former is the fraction of dark matter as PBH whereas the latter is related to the linear threshold overdensity.

To constrain the parameters which characterise extended PBH mass distributions it is necessary to take into account the range of validity of each constraint separately, as each observation is sensitive to the effects of PBHs of different masses.  Also, since the minimum and maximum mass of PBHs within the horizon, changes with redshift, this should also be taken into account.  We follow \citet{Sureda:2021} in order to be able to adapt the monochromatic constraints from the literature to our extended mass functions with a mass spike \citep[see also][]{Bellomo_2018}.

To constrain the fraction of dark matter in PBHs, $f_{\text{PBH}}$, we also consider the same observational constraints as in \cite{Sureda:2021} excluding disputed constraints.

Here we focus our analysis on three representative characteristic masses $M_{*}= \{1\times10^{-10}, 50, 100\}M_{\odot}$ and set the blue index  to $n_{b}=3.5$ and $2.0$ for HC and FCT, respectively.  These values lie within the ranges that  do not always allow 100\% of DM in PBHs in extended mass functions of \citet{Sureda:2021}, but that with the adequate spike, manage to reach $f_{\text{PBH}}=1$.  For each $M_{*}$ we consider the following positions for the peak mass $M_{p}=\{10^{-4},10^{-3},10^{-2},10^{-1},0.5,1,10,100\} \times M_{*}$. 

We calculate the amplitude $A_{2}$ that allows to obtain the highest fraction of mass in PBHs with $M_{\text{p}}$. With these $A_{2}$ values we measure the highest value of $f_{\text{PBH}}$ from the set of observational constraints of \cite{Sureda:2021}.  We do this to study the dependence on the position of the spike in terms of the values of $\delta_c$, of the ratio $\delta_c/\left< |\delta|^{2} \right>^{1/2}$, and of the allowed fraction of DM that could be stored in PBHs given the set of constraints  adopted here.

The ratio $\delta_c/\left< |\delta|^{2} \right>^{1/2}$ can be calculated assuming all of the volume of each patch with an overdensity above $\delta_c$ collapses to form PBHs, or that a fraction of them do so, as encoded by $f_m$.  Therefore we do the calculation assuming the two extremes, $f_m=1$ and $f_m=\beta(a)=\rho_{\rm matter}(a)/\rho_{\rm relat}(a)$, the ratio of matter to relativistic energy densities at scale factor $a$.  This ratio also depends on whether we average the typical overdensities over all masses, or only on the masses surrounding the peak.  This is an important detail as we are interested in the meaningful use of the Press Schechter formalism for the larger fraction of PBHs in a sample.

Figure \ref{fig:HC} shows the resulting values of $\delta_c$ (left), the ratio of typical to collapse overdensities (middle), and the upper limit on the fraction of DM in PBHs (right panels) for the HC scenario for three characteristic masses, $10^{-10}, 50$, and $100M_\odot$ (top to bottom, respectively), as a function of spike mass.  Different colours in the left and center columns correspond to the two choices of $f_m$.  All panels show the unit value as a horizontal line.  In all cases it can be seen that for $M_{p}\sim M_*$ it is possible to obtain collapse threshold overdensities $\delta_c\sim 1$, consistent with estimates from numerical relativistic simulations \citep{Niemeyer:1998PhRvL,Niemeyer:PRD99}, particularly for $f_m=1$. The figure shows that at this mass scale there is a sharp transition from $\delta_c \ll 1$ to $\delta_c \gg 1$. Mass functions with spikes whose mass scale $M_p$ is larger than the characteristic mass $M_*$ are almost monochromatic at $M_p$, having an extended component with a small amplitude. Therefore, in order to have an important fraction of mass beyond $M_*$, the required $\delta_c$ for collapse should be large to avoid collapse on scales other than that of the spike in the power spectrum, explaining the jump in $\delta_c$. 
In any case, it is possible to obtain ratios of critical to typical overdensities of order $\sim 1$ for the same values of spike mass but mostly for the $f_m=\beta$ case, when averaging typical overdensities over all PBH masses.  We checked that by lowering the width of the log-normal spike, it is also possible to increase this ratio to $\sim 1$ for the $f_m=1$ case. 

The right column shows that 
the fraction of DM in PBHs can reach values of $f_{\rm PBH}>1$, that also exceed the one for the extended case without the spike (shown as a yellow hoizontal line) when $M_p\lesssim M_*$, i.e., there are instances where the spike can make the combination of spike and extended distributions of better candidates to conform all of the dark matter.  The  colour bar on the right in this column shows the fraction of the mass density stored in the spike of the mass function, which is $\sim 50\%$ for the cases with $f_{\text{PBH}}\geq1$. This is only shown  for $f_m=1$.

Figure \ref{fig:FCT} shows the results for the FCT scenario for the same three characteristic masses, as a function of spike mass. As can be seen, in this scenario we reproduce the result of the HC case, that the fractions of DM in PBH show a remarkable increase when a spike mass similar to the characteristic one is adopted, particularly for low values of $M_*$.

We stress the fact that the average overdensities in the spike of the power spectrum depend on the width of the spike $\epsilon$, which can be understood as follows. As it was explained earlier, when $M_{p}\simeq M_{*}$, the average overdensity is dominated by the spike. Then, as the typical fluctuation is related to the power spectrum at the typical scale and, as for small width the height in power of the spike is large, it follows that the typical fluctuation will be sensitive to the width $\epsilon$ when $M_{p}\simeq M_{*}$. One can therefore modify this parameter accordingly to make the average overdensities in the spike comparable to, or much larger than, $\delta_c$.   It is indeed possible to reconcile a unit ratio for $\delta_c/\left< |\delta|^{2} \right>^{1/2}|_{\text{peak}}$ for the $f_m=1$ cases shown in both figures \ref{fig:HC} and \ref{fig:FCT} by lowering the width $\epsilon$ by enough orders of magnitude.  This  shows that for featureless spikes, such as a Dirac delta function, which can be approximated by our log-normal function with $\epsilon \to 0$, simply all overdense patches of the scale of the spike collapse, making the details of the distribution function of fluctuations unimportant for the abundance of PBHs of this mass.  However, it is not guaranteed that unit values for $\delta_c$ and  $\delta_c/\left< |\delta|^{2} \right>^{1/2}$ would result for Press-Schechter PBH mass functions for  specific models such as, for example, double field inflation \citep{Palma:2020ejf}, where the shape and width of the spike is fixed by the model itself.

As mentioned above, the collapse overdensity is the expected one, $\sim1$, for models where the spike and characteristic masses are similar, $M_p\sim M_*$, and can be made to satisfy the condition of similar critical and typical overdensities for $f_m=1$.  These particular mass functions from power spectra with spikes hold a strong connection with the primordial power spectrum via the PS formalism and they are quite similar to monochromatic mass functions, although containing also populations of PBHs of masses above or below the mass of the spike, including relics, if these turn out to be stable.

\section{Conclusions} \label{sec:conclusions}

The possibility of production of primordial black holes in the early universe covering a wide range of masses, has regained new motivation by the detection of gravitational waves by LIGO/Virgo. In this paper we extend the \citet{Sureda:2021} formalism, which allows to construct a mass function of PBHs from primordial power spectra with blue indices, to include an additional log-normal spike. The resulting mass distributions include both an extended component and a monochromatic component. 

The PBH mass functions thus obtained 
depend on extra parameters such as the location, width and amplitude of the spike in the primordial power spectrum, which are added to the pivot scale and value of the blue index for a broken spectrum. In order to reduce the dimensionality of the problem, we concentrate on the study of extended mass functions with exponential breaks in $M^*=10^{-10}, 50$ and $100M_\odot$, choose an amplitude for the spike such that it maximises the mass density of PBHs in the spike, and fix the width of the spike to $\epsilon=0.01$.  We let the wavenumber or, equivalently, mass of the spike to vary.  We find that for these particular choices, which in \citet{Sureda:2021} are found only in some cases to be able to conform all of the DM, there are values of the spike mass that are quite interesting and always allow $f_{\text{PBH}}=1$.

When the spike mass is similar to either of these three characteristic mass values we find that the critical overdensity for collapse can be $\sim 1$ in agreement with the values preferred by fully relativistic simulations. 
Furthermore, we find that with the right choice of spike width, $\epsilon$, the typical overdensity values can be similar to (or much larger for width $\epsilon \to 0$ than) the collapse threshold (both conditions are needed for a meaningful use of the Press-Schechter formalism), implying that even though our power spectra may very well be non-linear, particularly on the scale of the spike, our approximation would be largely free of effects from non-gaussianities in the distribution of fluctuations.  Notice that this simplification is particularly valid for extended mass functions with a spike, as a large fraction of the mass is in the spike itself.  For extended mass distributions without a spike, it would still be important to have a better understanding of the distribution function of overdensities, as deviations of gaussianity could indeed be important to properly infer the shape of the high mass tail of the distribution. 

We find that the HC and FCT cases with $M_p=50M_\odot$ and $M_*=100M_\odot$ not only represent PBH mass functions that could compose all of the DM according to current constraints, but these could also provide $\sim 1000M_\odot$ seeds for the supermassive black holes at the centres of galaxies dominated by active galactic nucleus feedback (e.g. \citealt{Lagos,Ferrara}): the abundance of PBH with masses $M_{\rm PBH}> 1940$ and $940 M_\odot$, for HC and FCT respectively, match the abundance of the host dark matter haloes of AGN feedback dominated galaxies, that is, the abundance of haloes with $M_{\rm halo}>3\times10^{12}M_\odot$ \citep{HernandezAguayo}. Notice that PBHs of masses similar to either the characteristic mass $M^{*}$ or the spike mass $M_{p}$ do not affect the $1000M_{\odot}$ SMBH seeds through any direct physical mechanism, but rather, the resulting mass function is able to account for the number of the aforementioned seeds.

Another interesting result is that if PBHs do not evaporate completely, leaving behind a stable relic instead (see for instance \citealt{Araya:2022prep}), then these could outnumber the PBHs in the spike by up to $15$ orders of magnitude.  Additionally, the population of PBHs in the power law region of the extended mass function, can also be as numerous, or even more so, than the PBHs in the monochromatic spike.  This is valid for the mass function parameters that allow all of the DM in PBHs.  Therefore, even though the abundance of the PBHs in the spike may make them difficult to detect, the mass functions explored here show that, in addition to the almost monochromatic PBHs of the spike, there could be a much larger population of PBHs of different masses and, possibly, relic PBHs in the Universe.

\section*{Acknowledgements}

We thank the anonymous referee for thoughtful remarks and suggestions. We thank Sebasti\`en Clesse and Federico Stasyszyn for helpful discussions. 
This project has received funding from the European Union's Horizon 2020 Research and Innovation Programme under the Marie Sk\l{}odowska-Curie grant agreement No 734374. JM acknowledges the support from ANID REDES 190147. MR acknowledges financial support provided under the European Union’s H2020 ERC Consolidator Grant “Gravity from Astrophysical to Microscopic Scales”, grant agreement no. GRAMS-815673 and INFN (Istituto Nazionale di Fisica Nucleare). IA acknowledges funding from ANID, REC Convocatoria Nacional Subvenci\'on a Instalaci\'on en la Academia Convocatoria A\~no 2020, Folio PAI77200097. NP acknowledges support from a RAICES grant from the Ministerio de Ciencia, Tecnología e Innovación, Argentina.

\section*{Data Availability}
No new data were generated or analysed in support of this
research.



\bibliographystyle{mnras}
\bibliography{biblio}
{}




\appendix

\section{Variance and Mass function calculations}\label{sec:mathematical}

In this appendix we explicit some of the required calculations throughout this work. 

The square of the variance $\sigma^2(M)$ is computed as

\begin{widetext}

\begin{eqnarray}
\sigma^{2}(M)&=& 4\pi A_{1} a_{\text{m}}^{4} \left[
\int_0^{k_{\text{piv}}}{k^{n_s} k^{2}\,\text{d}k} + \int_{k_{\text{piv}}}^{k_R}{
k_{piv}^{n_s-n_b}k^{n_b} k^{2}\,\text{d}k}+ \frac{A_2}{A_{1}}\frac{1}{\sqrt{2\pi} \epsilon} \int_0^{k_{R}}{k^2e^{-\frac{\log^2{(k/k_c)}}{2\epsilon^{2}}}\text{d}k} \right] \nonumber \\
 &=& 4\pi\,A_1\, a_{\text{m}}^{4}\,  \left[\frac{k^{n_s+3}}{n_s+3}\Bigg \vert_{0}^{k_{\text{piv}}}+\frac{k_{\text{piv}}^{n_s-n_b}k^{n_b+3}}{n_b+3}\Bigg \vert_{k_{\text{piv}}}^{k_R} -  \frac{A_2}{A_1}\frac{\epsilon k_{c}^{3}}{2} e^{\frac{9\epsilon^2}{2}} \left( \mathrm{Erf}\left(\frac{\log{(k_c/k)}+3\epsilon^{2}}{\sqrt{2}\epsilon}\right)\right)\Bigg \vert^{k_{R}}_{0}\right] \nonumber\\
 &=& 4\pi\,A_1\, a_{\text{m}}^{4}\,  \left[\frac{k_{\text{piv}}^{n_s+3}}{n_s+3} + \frac{k_{\text{piv}}^{n_s-n_b}k_{R}^{n_b+3}}{n_b+3}+\frac{k_{\text{piv}}^{n_s-n_b}k_{\text{piv}}^{n_b+3}}{n_b+3}+ \frac{A_2}{2 A_1}  k_{c}^{3} e^{\frac{9\epsilon^2}{2}} \left[1- \mathrm{Erf}\left(\frac{\log{(k_c/k_R)}+3\epsilon^{2}}{\sqrt{2}\epsilon}\right)\right]\right],
\end{eqnarray}
where $a_{\text{m}}$ corresponds to the model scale factor (HC or FCT) and $\text{Erf}$ is the error function, given by
\begin{equation}
    \text{Erf}\,(x)=\frac{2}{\sqrt{\pi}}\int_0^{x}{e^{-t^2}\,\text{d}t}\nonumber.
\end{equation}
By rearranging, we obtain
\begin{equation}\label{eq: computed sigma squared3}
    \sigma^2(M) = \frac{4\pi\,A_1\, a_{\text{m}}^{4} k_{\text{piv}}^{n_s-n_b}}{(n_s+3)(n_b+3)}\, \left[(n_b-n_s)k_{\text{piv}}^{n_b+3} + (n_s+3)k_{R}^{n_b+3} + 
    \frac{A_2}{2 A_1}\frac{(n_s+3)(n_b+3)}{k_{\text{piv}}^{n_s-n_b}}  k_{c}^{3} e^{\frac{9\epsilon^2}{2}} \left[1- \mathrm{Erf}\left(\frac{\log{(k_c/k_R)}+3\epsilon^{2}}{\sqrt{2}\epsilon}\right)\right]\right].
\end{equation}

\subsection{HC scenario}
In this scenario there is a relation between the PBH mass and the scale factor, namely,
\begin{equation}
a_{\text{HC}}=\frac{C_{\text{HC}}G}{\pi c^{2} }\frac{\sqrt{M}}{f_m}.
\end{equation}
Also, the wavenumber is related to the PBH mass as
\begin{equation}
    k_{\text{HC}}= \frac{C_{\text{HC}}}{\sqrt{M}},
\end{equation}
for which the variance reads
\begin{eqnarray}
    \sigma^2(M) &=& \frac{4\pi\,A_1\, C_{\text{HC}}^{(n_s-nb)}\,M_{\text{piv}}^{-(n_s-n_b)/2}}{(n_s+3)(n_b+3)} \left(\frac{G C_{\text{HC}}}{\pi c^2 f_m}\right)^4\left[(n_b-n_s)C_{\text{HC}}^{n_b+3} M_{\text{piv}}^{-\frac{n_b+3}{2}} M^{2} + (n_s+3)C_{\text{HC}}^{n_b+3}M^{\frac{1-n_b}{2}} +\right.\nonumber\\
    &&\left.\frac{A_2}{2 A_1}\frac{(n_s+3)(n_b+3)}{C_{\text{HC}}^{n_s-nb} M_{\text{piv}}^{-(n_s-n_b)/2}} C_{\text{HC}}^{3} M_{\text{p}}^{-3/2} e^{\frac{9\epsilon^2}{2}} \left[1-\mathrm{Erf}\left(\frac{\log{\sqrt{M/M_{\text{p}}}}+3\epsilon^{2}}{\sqrt{2}\epsilon}\right)\right]\right]. 
    \label{eq:sigmaM_hc}
\end{eqnarray}
We can write $\sigma(M)=\frac{4\pi\,A_1\, C_{\text{HC}}^{(n_s-nb)}\,M_{\text{piv}}^{-(n_s-n_b)/2}}{(n_s+3)(n_b+3)} \left(\frac{G C_{\text{HC}}}{\pi c^2 f_m}\right)^4 F(M)$, where the function $F(M)$ is defined as
\begin{eqnarray}
F(M)&=&\left[(n_b-n_s)C_{\text{HC}}^{n_b+3} M_{\text{piv}}^{-\frac{n_b+3}{2}} M^{2} + (n_s+3)C_{\text{HC}}^{n_b+3}M^{\frac{1-n_b}{2}} +\right.\nonumber\\
    &&\left.\frac{A_2}{2 A_1}\frac{(n_s+3)(n_b+3)}{C_{\text{HC}}^{n_s-nb} M_{\text{piv}}^{-(n_s-n_b)/2}} C_{\text{HC}}^{3} M_{\text{p}}^{-3/2} e^{\frac{9\epsilon^2}{2}} \left[1-\mathrm{Erf}\left(\frac{\log{\sqrt{M/M_{\text{p}}}}+3\epsilon^{2}}{\sqrt{2}\epsilon}\right)\right]\right]. 
\end{eqnarray}

Now, recalling the function $\nu(M)$ given by
\begin{equation}
    \nu=\frac{\delta_{c}}{\sigma(M)},
\end{equation}
we find it useful to introduce the critical value $\nu_*$, defined by
\begin{equation}
    \nu_{*}:=\nu(M_*).
\end{equation}
Then, the critical mass $M_*$ is defined such that $\nu_*=1$, which directly gives a way to compute $\delta_c$ from the variance, namely
\begin{equation}
    \delta_c=\sigma(M_*).
\end{equation}

\subsection{FCT scenario}

For this scenario, we get the following relation for the wavenumber:
\begin{equation}
    k_{\text{FCT}} = \frac{C_{\text{FCT}}}{\sqrt[3]{M}}.
\end{equation}
Then, we have that
\begin{eqnarray}
    \sigma^2(M) &=& \frac{4\pi\,A_1\, a_{\text{FCT}}^{4} C_{\text{FCT}}^{(n_s-n_b)}M_{\text{piv}}^{-(n_s-n_b)/3}}{(n_s+3)(n_b+3)}\, \left[(n_b-n_s)C_{\text{FCT}}^{n_b+3} M_{\text{piv}}^{-\frac{n_b+3}{3}} + (n_s+3)C_{\text{FCT}}^{n_b+3}M^{-\frac{n_b+3}{3}} + \right.\nonumber\\
    &&\left. \frac{A_2}{2 A_1}\frac{(n_s+3)(n_b+3)}{M_{\text{piv}}^{-(n_s-n_b)/3}} C_{\text{FCT}}^{3} M_{\text{p}}^{-1} e^{\frac{9\epsilon^2}{2}} \left[1- \mathrm{Erf}\left(\frac{\log{\sqrt[3]{M/M_{\text{p}}}}+3\epsilon^{2}}{\sqrt{2}\epsilon}\right)\right]\right]. \nonumber\\
    \label{eq:sigmaM}
\end{eqnarray}
We can write
$
\sigma(M)= \left(\frac{4\pi\,A_1\, a_{\text{FCT}}^{4} C_{\text{FCT}}^{(n_s-n_b)}M_{\text{piv}}^{-(n_s-n_b)/3}}{(n_s+3)(n_b+3)}\right)^{1/2}F(M) 
$,
where the function $F(M)$ is defined as
\begin{eqnarray}
F(M)&=&\left[(n_b-n_s)C_{\text{FCT}}^{n_b+3} M_{\text{piv}}^{-\frac{n_b+3}{3}} + (n_s+3)C_{\text{FCT}}^{n_b+3}M^{-\frac{n_b+3}{3}} + \right.\nonumber\\
    &&\left. \frac{A_2}{2 A_1}\frac{(n_s+3)(n_b+3)}{M_{\text{piv}}^{-(n_s-n_b)/3}} C_{\text{FCT}}^{3} M_{\text{p}}^{-1} e^{\frac{9\epsilon^2}{2}} \left[1- \mathrm{Erf}\left(\frac{\log{\sqrt[3]{M/M_{\text{p}}}}+3\epsilon^{2}}{\sqrt{2}\epsilon}\right)\right]\right]^{1/2}. \nonumber\\
    \label{eq:gM}  
\end{eqnarray}
Now, in order to get the analytic expression for the spiky mass function, we need to evaluate the derivative of $\nu$ with respect to $M$. Indeed, we get,
\begin{eqnarray}\label{eq:dnudm}
\frac{\text{d}\nu}{\text{d}M}&=&\frac{1}{6}\frac{F(M_{*})}{F(M)^{3}}
\left[(n_b+3)(n_s+3)C_{\text{FCT}}^{n_b+3} M^{-\frac{n_b+6}{3}}+\right. \nonumber\\
&& \left. \frac{A_2}{A_1\sqrt{2\pi}\epsilon}\frac{(n_s+3)(n_b+3)}{M_{\text{piv}}^{-(n_s-n_b)/3}} C_{\text{FCT}}^{3}  \frac{1}{M_{\text{p}}\,M} e^{\frac{9\epsilon^2}{2}} e^-{\left(\frac{\log{\sqrt[3]{M/M_{\text{p}}}}+3\epsilon^{2}}{\sqrt{2}\epsilon}\right)^2}
\right],
\end{eqnarray}

Finally, the mass function in this scenario is given by
\begin{eqnarray}
\frac{dn}{dM}&=&\frac{\rho_{\text{DM}}}{3\sqrt{2\pi}}\frac{F(M_{*})}{\, F(M)^{3}} \left[(n_b+3)(n_s+3)C_{\text{FCT}}^{n_b+3} M^{-\frac{n_b+9}{3}}+
\right.\nonumber\\
&&\left.\frac{A_2}{ A_1 \sqrt{2\pi}\epsilon}\frac{(n_s+3)(n_b+3)}{M_{\text{piv}}^{-(n_s-n_b)/3}}\frac{C_{\text{FCT}}^{3}}{M_{\text{p}}\,M^2} e^{\frac{9\epsilon^2}{2}} e^-{\left(\frac{\log{\sqrt[3]{M/M_{\text{p}}}}+3\epsilon^{2}}{\sqrt{2}\epsilon}\right)^2}
\right] e^{-\frac{1}{2}\left(\frac{\nu_{*}}{\nu_{m}}\right)^2 }.\nonumber\\
\end{eqnarray}

\end{widetext}

\label{lastpage}

\bsp	
\label{lastpage}
\end{document}